\documentclass[12pt]{article}
\usepackage{epsfig}
\usepackage{amsmath}
\usepackage{hhline}
\usepackage{amssymb}
\usepackage{times}
\usepackage{cite}
\usepackage{lineno}

\newlength{\dinwidth}
\newlength{\dinmargin}
\begin{document}  
\newcommand{\pom}{{I\!\!P}}
\newcommand{\reg}{{I\!\!R}}
\def\gsim{\,\lower.25ex\hbox{$\scriptstyle\sim$}\kern-1.30ex%
\raise 0.55ex\hbox{$\scriptstyle >$}\,}
\def\lsim{\,\lower.25ex\hbox{$\scriptstyle\sim$}\kern-1.30ex%
\raise 0.55ex\hbox{$\scriptstyle <$}\,}
\newcommand{\trm}{m_{\perp}}
\newcommand{\trp}{p_{\perp}}
\newcommand{\trmm}{m_{\perp}^2}
\newcommand{\trpp}{p_{\perp}^2}
\newcommand{\alp}{\alpha_s}
\newcommand{\alps}{\alpha_s}
\newcommand{\sqrts}{$\sqrt{s}$}
\newcommand{\LO}{$O(\alpha_s^0)$}
\newcommand{\Oa}{$O(\alpha_s)$}
\newcommand{\Oaa}{$O(\alpha_s^2)$}
\newcommand{\PT}{p_{\perp}}
\newcommand{\JPSI}{J/\psi}
\newcommand{\PO}{I\!\!P}
\newcommand{\xbj}{x}
\newcommand{\xpom}{x_{\PO}}
\newcommand{\dgr}{^\circ}
\newcommand{\gev}{\,\mbox{GeV}}
\newcommand{\GeV}{\rm GeV}
\newcommand{\xp}{x_p}
\newcommand{\xpi}{x_\pi}
\newcommand{\xg}{x_\gamma}
\newcommand{\xgj}{x_\gamma^{jet}}
\newcommand{\xpj}{x_p^{jet}}
\newcommand{\xpij}{x_\pi^{jet}}
\renewcommand{\deg}{^\circ}
\newcommand{\qsq}{\ensuremath{Q^2} }
\newcommand{\gevsq}{\ensuremath{\mathrm{GeV}^2} }
\newcommand{\et}{\ensuremath{E_t^*} }
\newcommand{\rap}{\ensuremath{\eta^*} }
\newcommand{\gp}{\ensuremath{\gamma^*}p }
\newcommand{\dsiget}{\ensuremath{{\rm d}\sigma_{ep}/{\rm d}E_t^*} }
\newcommand{\dsigrap}{\ensuremath{{\rm d}\sigma_{ep}/{\rm d}\eta^*} }
\newcommand {\gapprox}
   {\, \raisebox{-0.7ex}{$\stackrel {\textstyle>}{\sim} \,$}}
\newcommand {\lapprox}
   {\, \raisebox{-0.7ex}{$\stackrel {\textstyle<}{\sim} \,$}}

\def\Journal#1#2#3#4{{#1} {\bf #2}, #4 (#3)}
\def\NCA{\em Nuovo Cimento}
\def\NIM{\em Nucl. Instrum. Methods}
\def\NIMA{{\em Nucl. Instrum. Methods} {\bf A}}
\def\NPB{{\em Nucl. Phys.}   {\bf B}}
\def\PLB{{\em Phys. Lett.}   {\bf B}}
\def\PRL{\em Phys. Rev. Lett.}
\def\PRD{{\em Phys. Rev.}    {\bf D}}
\def\PR{{\em Phys. Rev.}    }
\def\PRP{{\em Phys. Rep.}    }
\def\ZPC{{\em Z. Phys.}      {\bf C}}
\def\ZP{{\em Z. Phys.}      }
\def\EJC{{\em Eur. Phys. J.} {\bf C}}
\def\EJA{{\em Eur. Phys. J.} {\bf A}}
\def\CPC{\em Comp. Phys. Commun.}
\def\SJNP{{\em Sov. J. Nucl. Phys.}}
\def\SPJETP{{\em Sov. Phys. JETP}}
\def\JETPL{{\em JETP Lett.}}


\begin{titlepage}

\noindent
\begin{flushleft}
DESY 06-048\hfill ISSN 0418-9833\\
May 2006
\end{flushleft}

\vspace*{2cm}

\begin{center}
\begin{Large}

{\boldmath \bf      
    Diffractive Deep-Inelastic Scattering \\
with a Leading Proton at HERA
}

\vspace{2cm}

H1 Collaboration

\end{Large}
\end{center}

\vspace{2cm}

\begin{abstract}
\noindent
The cross section
for the diffractive deep-inelastic 
scattering process $ep \rightarrow e X p$ is measured, 
with the leading final state 
proton detected in the H1 Forward Proton Spectrometer.
The data analysed cover the range 
$\xpom <0.1$ in fractional proton longitudinal momentum loss,
$0.08 < |t| < 0.5 \ {\rm GeV^{-2}}$ in squared four-momentum transfer at the 
proton vertex, 
$2< Q^2 <50~{\GeV^2}$ in photon virtuality
and $0.004 < \beta = x / \xpom < 1$, where $x$ is the Bjorken scaling
variable. 
For $\xpom \lapprox 10^{-2}$, 
the differential cross 
section has a dependence of approximately
${\rm d} \sigma / {\rm d} t \propto e^{6 t}$, independently 
of $\xpom$, $\beta$ and $Q^2$ within uncertainties. 
The cross section is also 
measured triple differentially in $\xpom$, $\beta$ and 
$Q^2$. The $\xpom$ dependence is interpreted 
in terms of an effective pomeron trajectory with intercept 
$\alpha_{\pom}(0)=1.114 \pm 0.018 \ ({\rm stat.}) 
\pm 0.012 \ ({\rm syst.}) \
^{+0.040}_{-0.020} \ ({\rm model})$ 
and a sub-leading exchange. The data are
in good agreement with an H1 measurement for which 
the event selection is based on
a large gap in the rapidity distribution of the
final state hadrons, after accounting for proton 
dissociation contributions in the latter.
Within uncertainties, the dependence of the
cross section on $x$ and $Q^2$
can thus be factorised from the dependences on all studied variables
which characterise the proton vertex, for both the pomeron and
the sub-leading exchange.
\end{abstract}

\vspace{1cm}

\begin{center}
Submitted to {\it Eur. Phys. J.} {\bf C}
\end{center}

\end{titlepage}

\begin{flushleft}

A.~Aktas$^{9}$,                
V.~Andreev$^{25}$,             
T.~Anthonis$^{3}$,             
B.~Antunovic$^{26}$,           
S.~Aplin$^{9}$,                
A.~Asmone$^{33}$,              
A.~Astvatsatourov$^{3}$,       
A.~Babaev$^{24, \dagger}$,     
S.~Backovic$^{30}$,            
A.~Baghdasaryan$^{37}$
% ...

\bigskip{\it
% ...
$ ^{3}$ Inter-University Institute for High Energies ULB-VUB, Brussels;
          Universiteit Antwerpen, Antwerpen; Belgium$^{ c}$ \\
% ...
$ ^{9}$ DESY, Hamburg, Germany \\
% ...
$ ^{24}$ Institute for Theoretical and Experimental Physics,
          Moscow, Russia$^{ k}$ \\
 $ ^{25}$ Lebedev Physical Institute, Moscow, Russia$^{ e}$ \\
 $ ^{26}$ Max-Planck-Institut f\"ur Physik, M\"unchen, Germany \\
% ...
$ ^{30}$ Faculty of Science, University of Montenegro,
          Podgorica, Serbia and Montenegro$^{ e}$ \\
% ...
$ ^{33}$ Dipartimento di Fisica Universit\`a di Roma Tre
          and INFN Roma~3, Roma, Italy \\
% ...
$ ^{37}$ Yerevan Physics Institute, Yerevan, Armenia \\
% ...
$ ^{\dagger}$ Deceased \\

\bigskip
% ...
$ ^c$ Supported by FNRS-FWO-Vlaanderen, IISN-IIKW and IWT
      and  by Interuniversity
Attraction Poles Programme,
      Belgian Science Policy \\
% ...
$ ^e$ Supported by the Deutsche Forschungsgemeinschaft \\
% ...
$ ^k$ Partially Supported by Russian Foundation
      for Basic Research,  grants  03-02-17291
      and  04-02-16445 \\
}

\end{flushleft}

\newpage

\section{Introduction}

Diffractive processes such as $ep \rightarrow eXp$ have been
studied  extensively in deep-inelastic electron\footnote{For simplicity, 
the incident and scattered leptons are always referred to in the following
as `electrons', although the data studied here were obtained with 
both electron and positron beams.}-proton
scattering (DIS) at the HERA 
collider~\cite{H1Diff,H1Diff94,ZEUSDiff,ZEUSMX,H1LRG,H1FPS1,H1FPS2,ZEUSLPS},
since understanding them in detail is fundamental 
to the development of 
quantum chromodynamics (QCD) at high parton densities. 
The photon virtuality $Q^2$ supplies a hard scale for the application of 
perturbative QCD, so that diffractive 
DIS events can be viewed as processes in which 
the photon probes a net colour singlet 
combination of exchanged partons.
A hard scattering QCD collinear factorisation theorem \cite{Collins} 
allows `diffractive parton distribution functions' (DPDFs) to be defined,
expressing proton parton probability
distribution functions under the condition of a particular scattered proton
four-momentum. The $x$ and $Q^2$ dependences of diffractive DIS can
thus be treated
with a similar theoretical description to that applied to 
inclusive DIS, for example through the application of the
DGLAP parton evolution equations~\cite{DGLAP}.

\begin{figure}[h] \unitlength 1mm
 \begin{center}
 \begin{picture}(100,60)
  \put(25,-6){\epsfig{file=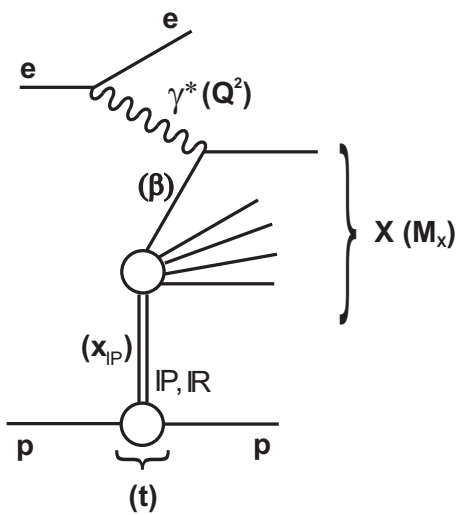,width=0.375\textwidth}}
 \end{picture}
 \end{center}
 \caption{Schematic illustration of the diffractive
DIS process $ep \rightarrow eXp$ and the kinematic variables used for
its description in a model in which the pomeron ($\pom$) and a sub-leading
($\reg$) trajectory are exchanged.}
\label{reggefac}
\end{figure}

Within Regge phenomenology, 
diffractive cross sections are described 
by the exchange of a leading pomeron ($\pom$)
trajectory, as illustrated in figure~\ref{reggefac}.
H1 diffractive DIS
data \cite{H1LRG} have been interpreted in a combined framework, which applies
the QCD factorisation theorem to the $x$ and $Q^2$ dependences and
uses a Regge inspired approach to express the dependence on the 
fraction $\xpom$ of the incident proton longitudinal
momentum carried by the colour singlet exchange.
The data at low $\xpom$ are well described in this framework and
DPDFs and a pomeron trajectory intercept 
have been extracted. In order to describe the data at
larger $\xpom$, it is necessary to include
a sub-leading exchange trajectory ($\reg$), with an intercept which is
consistent \cite{H1Diff94}
with the approximately degenerate trajectories associated
with the $\rho$, $\omega$, $a_2$ and $f_2$ mesons. 

In many previous analyses, including \cite{H1LRG}, diffractive DIS events
are selected on the basis 
of the presence of  
a large rapidity gap (LRG) between the leading 
proton and the remainder $X$ of the hadronic final state.
A complementary way to study diffractive processes is by direct 
measurement of the outgoing proton using the H1 
Forward Proton Spectrometer (FPS)~\cite{FPS,H1FPS1} or its
ZEUS counterpart~\cite{ZEUSLPS}.
Although the available statistics are smaller, the FPS method of studying
diffraction has several advantages over the LRG method.  
In contrast to the LRG case, the 
squared four-momentum transfer at the proton vertex $t$ 
can be reconstructed.
The FPS method also selects 
events in which the proton scatters elastically, whereas
the LRG method does not distinguish the elastic case
from dissociation to
excited systems $Y$ with small masses $M_Y$.
The FPS also allows measurements up to 
higher values of $\xpom$ 
than is possible with the LRG method, extending into regions
where the sub-leading trajectory is the dominant exchange. Together,
the FPS and LRG data thus provide a means of testing in detail 
the extent to which
the variables $\xpom$, $t$ and $M_Y$ associated with the proton vertex 
can be factorised from the variables 
$\beta = x / \xpom$ and $Q^2$ describing the
hard interaction.

In this paper, a measurement of the cross section for the diffractive
DIS process $ep \rightarrow e X p$ using the FPS
is reported.
The $t$ dependence is presented in the form of a differential cross section 
$\xpom \ {\rm d}^2 \sigma / {\rm d}t \, {\rm d}\xpom$, from which the
exponential slope of the $t$ distribution is measured
and its dependence on other variables studied. 
Diffractive reduced cross sections, $\sigma_r^{D(4)}(\beta,Q^2,\xpom,t)$ 
at $|t|=0.25~\GeV^2$, and 
$\sigma_r^{D(3)}(\beta,Q^2,\xpom)$ integrated over $t$,
are also measured. 
These observables are used to investigate the dependences on $\beta$ and $Q^2$,
to extract the pomeron trajectory intercept from the $\xpom$
dependence and to
quantify the sub-leading exchange contribution. The data are also compared
directly with the LRG measurement \cite{H1LRG} in order
to test the compatibility between the results 
obtained with the two measurement techniques and to 
quantify the proton dissociation contribution in the LRG data.


\section{Experimental Technique}

The data used in this analysis correspond to an integrated
luminosity of 28.4 $\rm pb^{-1}$ and
were collected with the 
H1 detector in the years 1999 and 2000. In 
these years the HERA collider was operated 
at electron and proton beam energies of $E_e=27.6~\GeV$ and $E_p=920~\GeV$, 
respectively, corresponding to an $ep$ centre of mass energy of 
$\sqrt{s} = 319 \ {\rm GeV}$.

\subsection{H1 detector}
\label{detector}

A detailed description of the H1 detector can be found elsewhere
\cite{h1detector}. Here, the components most
relevant for the present measurement are described briefly.

Scattered electrons with polar angles\footnote{In the right-handed coordinate 
system used, 
the origin is at the nominal interaction point, with
the $+z$ axis and the 
polar angle $\theta = 0$ in the direction of the outgoing proton beam
(the `forward' direction).
The $+x$ axis points towards the centre of HERA. 
Transverse momenta are measured with respect to the beam axis.} 
in the range $153\deg<\theta_e^\prime <177\deg$ are measured in 
a lead\,/\,scintillating-fibre calorimeter, the SpaCal \cite{SPACAL}.
The energy resolution is
$\sigma(E)/E\approx 7\%/\sqrt{E[\GeV]}\oplus 1\%$
and the energy scale uncertainty varies between $2.0\%$ at 
a scattered electron energy of $E_e^\prime = 11~\GeV$ 
and $0.5\%$ at $E_e^\prime =27.6~\GeV$ \cite{H1F2}.
A Backward Drift Chamber (BDC) in front of the SpaCal
is used to measure the electron polar angle with a precision 
of 0.5~mrad and to suppress background where
neutral particles fake the scattered electron signal.
The SpaCal also has a hadronic section, with an energy scale known 
to a precision of $7\%$.

The Central Tracking Detector (CTD), with  a polar angle coverage of
$20\deg<\theta<160\deg$, is used 
to reconstruct the interaction vertex and
to measure the momentum of
charged particles from the curvature of their trajectories 
in the 1.15~T field provided by a superconducting solenoid.
The finely segmented Liquid Argon (LAr) sampling calorimeter \cite{LAR}
surrounds the tracking system and covers the range in polar angle 
$4\deg<\theta<154\deg$.
Its total depth varies with $\theta$ between 4.5 and 8 interaction lengths. 
The absolute hadronic energy scale is known with a 
precision of $4\%$ for the measurements presented here.
The hadronic final state is reconstructed using an
energy flow algorithm which 
combines charged particles measured in the CTD with  information from
the SpaCal and LAr calorimeters \cite{fscomb}.

The luminosity is determined  with a precision of $1.5\%$ by 
detecting photons from the Bethe-Heitler process 
$ep\rightarrow ep\gamma$ in a crystal \v{C}erenkov calorimeter,
located at $z=-103$~m.

The energy and scattering angle of the leading proton are obtained
from track measurements in the FPS \cite{FPS,H1FPS1}. 
Protons scattered through small angles  
are deflected by the proton beam-line magnets into a system of 
detectors placed within the proton beam pipe inside movable 
stations, known as Roman Pots. Each Roman Pot station contains
four planes of five scintillating fibres, which together measure two
orthogonal coordinates in the $(x,y)$ plane. 
The stations used in this analysis 
approach the beam horizontally from outside 
the proton ring and
are positioned at  
$z = 64 \ {\rm m}$ and $z = 80 \ {\rm m}$. 
The detectors are sensitive to scattered protons which 
lose less than $10\%$ of their energy in the $ep$ 
interaction and which are scattered through angles $\lapprox 1 \ {\rm mrad}$.

For each event, the leading proton energy and
the proton scattering angles at the interaction point 
in the horizontal ($x-z$) and vertical ($y-z$) planes
are obtained by applying transfer 
functions derived from the beam optics to the
track parameters reconstructed in the FPS. 
The scattered proton energy is thus measured independently using
the information in the horizontal and vertical planes. By comparison
of these results, it is inferred that the 
energy resolution is around 6~GeV, independently of energy within the
measured range, and that the absolute energy scale uncertainty is 0.5~GeV.
The uncertainties in the reconstruction of the
transverse momentum components $p_x$ and $p_y$ are
quantified using a sample of elastic $e p \rightarrow e \rho^0 p$ 
photoproduction events with $\rho^0 \rightarrow \pi^+ \pi^-$ decays.
By comparing the FPS measurements with values reconstructed 
from the charged pions in the CTD, the
resolution of the FPS is determined to be 
$\sim 40 \ {\rm MeV}$ for $p_x$ 
and $\sim 100 \ {\rm MeV}$ for $p_y$, dominated by the transverse momentum 
spread of the proton beam at the interaction point. The corresponding
$t$-resolution varies over the measured range from
$0.04 \ {\rm GeV^2}$ at $|t| = 0.08 \ {\rm GeV^2}$
to $0.08 \ {\rm GeV^2}$ at $|t| = 0.5 \ {\rm GeV^2}$.
The uncertainties in the
transverse momentum measurements are $10 \ {\rm MeV}$ for
$p_x$ and $30 \ {\rm MeV}$ for $p_y$.
The $t$-dependence measured in the FPS for the $\rho^0$ 
sample \cite{Karsch} is in 
good agreement with published H1 data \cite{rho:gammap}.
For a leading proton which passes through 
both FPS stations, the average overall track reconstruction efficiency is 
$20 \pm 2\%$, corresponding to the product of the efficiencies
in the individual scintillating fibre planes. The uncertainty 
on this efficiency is evaluated
by varying the details of the reconstruction procedure, for example the 
number of fibres per plane which are 
required to register a track element.

\subsection{Event selection and kinematic reconstruction}
\label{recsec}

The events used in this analysis are triggered on the basis of a coincidence 
between a track in the FPS, 
an electromagnetic cluster in the SpaCal calorimeter 
and a charged particle track providing an interaction vertex in the CTD. 
The trigger efficiency varies with the kinematic variables studied and is 
around $85\%$ on average.

Several selection criteria are applied to the data in order to suppress
beam related backgrounds, 
background due to photoproduction processes
and events in which 
the incoming electron loses 
significant energy through QED 
radiation. The DIS selection criteria are summarised below.

\begin{itemize}
\item The reconstructed $z$ coordinate  of the event vertex is required to 
lie within $35 \ {\rm cm}$ 
\linebreak
$(\sim 3 \sigma)$
of the mean position. At least one track originating 
from the interaction vertex and reconstructed in the 
CTD is required to have a transverse momentum above 0.1~GeV. 

\item The variables characterising the scattered electron, 
$E_e^\prime$ and $\theta_e^\prime$, 
are determined from the 
SpaCal cluster, linked to a reconstructed charged particle
track in the BDC, and
the interaction vertex reconstructed in the CTD.
The electron candidate is required to satisfy the criteria
$155\deg<\theta^{\prime}_e<176.5\deg$ 
and $E_e^\prime > 11 \ {\rm GeV}$. 

\item The quantity $E - p_z$, 
computed from the energies and longitudinal momenta of
all reconstructed particles including the electron, is required to lie 
between  $35~\GeV$ and $70~\GeV$. Neglecting detector effects
and QED radiation,
this quantity is expected to be twice the
electron beam energy for 
neutral current DIS events.
\end{itemize}

\noindent
The following requirements are applied 
to the leading proton measured in the FPS.

\begin{itemize}
\item The measurement is restricted to the region where the FPS acceptance is
high by requiring that the transverse momenta in the horizontal 
and vertical projections 
lie in the ranges $-0.38 < p_x < -0.24~\GeV$ and
$|p_y| < 0.7~\GeV$, respectively, and that 
the fractional energy of the leading proton, $E_p^\prime / E_p$, 
be greater than 0.9.  

\item To suppress cases where a DIS event reconstructed in the central
detector coincides with background 
in the FPS, for example due to an off-momentum beam proton (beam halo), 
the quantity $E + p_z$, summed 
over all reconstructed 
particles including the leading proton, is required to be 
below $1900~\GeV$. Neglecting detector effects, this 
quantity is expected to be twice the
proton beam energy for neutral current DIS events.
\end{itemize}

The inclusive DIS 
kinematic variables, $Q^2$, $x$ and the inelasticity $y$, are reconstructed 
using the techniques introduced in \cite{H1Diff94}.    
In order to optimise the resolution throughout the measured $y$ range,
information is exploited from both the scattered electron and
the hadronic final state according to  
\begin{eqnarray}
  y = y_e^2+y_d(1-y_d) \ \ \ ; \ \ \ 
Q^2=\frac{4E_e^2 (1-y)}{\tan ^2(\theta_e^\prime/2)} \ \ \ ; \ \ \ 
x=\frac{Q^2}{sy} \ .
\end{eqnarray}
Here, $y_e$ and $y_d$ denote the values of $y$
obtained from the scattered electron only (`electron method') and
from the angles 
of the electron and the hadronic final state (`double angle 
method'), respectively~\cite{dameth}. 
The analysis is restricted to the 
region $2 < Q^2 < 50~\GeV^2$ and $0.02 < y < 0.6$.

With $q$, $P$ and $P^\prime$ denoting the four-vectors of the exchanged 
virtual photon
and the incoming and outgoing protons, respectively, 
further variables specific to diffractive DIS are defined as
\begin{eqnarray}
 \xpom =\frac{q \cdot (P-P^\prime)}{q \cdot P} \ \ \ ; \ \ \
 \beta =\frac{Q^2}{2q \cdot (P-P^\prime)} \ ,
\end{eqnarray}
such that 
$\beta$ can be interpreted as the fraction of the 
colourless exchange longitudinal momentum which is carried
by the struck quark. 
Two different methods are used to reconstruct
these variables.
In the `leading proton' method, 
$\xpom$ is reconstructed directly from the energy of the leading proton, 
such that
\begin{eqnarray}
 \xpom = 1 - E'_p/E_p \ \ \ ; \ \ \ \beta =\frac{x}{\xpom} \ .
\end{eqnarray}
In the `$X$-mass' method, the mass of the system $X$ is first 
obtained from the hadrons reconstructed in the central detector using
\begin{eqnarray}
 M_X^2=(E^2-p_x^2-p_y^2-p_z^2)_{\rm had} \cdot \frac{y}{y_h} \ ,
\end{eqnarray}
where the subscript `${\rm had}$' represents a sum over all hadronic final
state particles excluding the leading proton
and $y_h$ is the value of $y$ reconstructed using only the hadronic 
final state~\cite{hadmeth}. Including the factor $y/y_h$
leads to cancellations of many measurement inaccuracies.
The diffractive
variables are then reconstructed using 
\begin{eqnarray}
 \beta = \frac{Q^2}{Q^2+M_X^2} \ \ \ ; \ \ \ 
\xpom =\frac{x}{\beta} \ .
\end{eqnarray}
The results obtained with the leading proton and $X$-mass 
methods agree well in the low $\xpom$ range where both are 
applicable. The $X$-mass method is used for $\xpom < 0.006$ 
and the leading proton method is used for $\xpom > 0.006$,
the choice being made on the basis of which
method provides the better resolution.

The squared four-momentum transfer $t = (P-P^\prime)^2$ is 
reconstructed using the transverse momentum $p_t$ of 
the leading proton measured with the FPS and the best value of $\xpom$
as described above, such that
\begin{eqnarray}
 t = t_{\rm min} - \frac{p_t^2}{1-\xpom} \ \ \ ; \ \ \
t_{\rm min} = - \frac{\xpom^2 m_p^2}{1-\xpom} \ ,
\end{eqnarray}
where $|t_{\rm min}|$ is the minimum kinematically accessible value of $|t|$
and $m_p$ is the proton mass.
In the analysis, the 
reconstructed $|t|$ is required to lie in the range 
$0.08 < |t| < 0.5~\GeV^2$.
The final data sample contains about 3\,300 events.


\section{Monte Carlo simulation and corrections to the data}
\label{mc}

Monte Carlo simulations are used to correct the data for the effects of 
detector acceptances and inefficiencies, migrations 
between measurement intervals due to finite resolutions and
QED radiation. The reaction $ep \rightarrow eXp$ 
is simulated
using an implementation of the `saturation' 
model~\cite{Saturation} within the RAPGAP generator~\cite{RAPGAP}. 
Following 
hadronisation using the Lund string model~\cite{Lund} as implemented in 
JETSET~\cite{Jetset}, the response of the H1 detector is simulated in detail and the events 
are passed through the same analysis chain as is used for the data.
Weights are applied to the generated events  
so that the important kinematic variable 
distributions are well described throughout the region of the measurement.

The background from photoproduction processes, where 
the electron is scattered into the backward beampipe and a particle from the
hadronic final state fakes the electron signature in the SpaCal, 
is estimated using the PHOJET Monte Carlo model~\cite{PHOJET}. This background 
is negligible except at the highest $y$ values and
is $6\%$ at most.
The proton dissociation background, where the leading proton originates
from the decay of a higher mass state, is estimated using an implementation 
in RAPGAP of the dissociation model originally developed for the 
DIFFVM Monte Carlo generator~\cite{DIFFVM}. This background 
is negligible except at the highest $\xpom$ values, where it reaches $2.7\%$. 
Background also arises from random 
coincidences of DIS events causing activity in the central detector
with beam-halo protons giving a signal 
in the FPS. This contribution is estimated statistically by
combining DIS events (without the requirement of a track in the 
FPS) with beam-halo protons from randomly triggered events. 
Subtractions of up to $7\%$ are made as a function of the total reconstructed 
$E+p_z$ of the event.

Cross sections are obtained at the Born level, using 
RAPGAP interfaced to the program HERACLES~\cite{HERACLES} to correct for
QED radiative effects.
The data are presented at
fixed $Q^2$, $\beta$, $\xpom$ and $t$ values, with corrections applied
for the influence of the
finite bin sizes using a parameterisation of the `2006 DPDF Fit A' to the H1 
LRG data~\cite{H1LRG} for the $Q^2$, $\beta$ and $\xpom$ dependences
and the $t$ dependences measured in this analysis at each 
$\xpom$ value (see section~\ref{tdep}). 


\section{Systematic Uncertainties on the Measured Cross Sections}
\label{systs}

Systematic uncertainties are considered 
from the following sources.

\begin{itemize}
\item The uncertainties in the leading proton 
energy, its
transverse momentum in the horizontal projection and that in the vertical 
projection are 0.5~GeV, 10~MeV and 30~MeV, 
respectively (see section~\ref{detector}). 
The corresponding average uncertainties on the 
$\sigma_r^{D(3)}$ and $\sigma_r^{D(4)}$ measurements are 
$5.7\%$, $6.0\%$ and $3.3\%$.

\item The energy scale uncertainty of the SpaCal implies an error
of between $0.5\%$ and
$2.0\%$ (depending on the energy) on the $E'_e$ measurement,
which leads to an average systematic error of $3.0\%$ on the $\sigma_r^D$
data points.
Possible biases in the $\theta_e^\prime$ measurement
at the level of $\pm 0.5$~mrad lead to an average systematic error of
2.4\%. 

\item The systematic uncertainties arising from 
the hadronic final state 
reconstruction are determined by varying the
hadronic energy scales of the LAr and SpaCal 
calorimeters by $4\%$ and $7\%$, respectively, and the energy fraction 
carried by tracks by $3\%$. Each of these sources leads to an uncertainty
in the $\sigma_r^D$ measurements of typically 1.5\%.

\item The model dependence of the acceptance and 
migration corrections is estimated by varying the shapes of the 
distributions in the kinematic variables $\xpom$, $\beta$ and $t$ 
in the RAPGAP simulation within the limits imposed by the 
present 
data. The $\xpom$ distribution is reweighted by $(1/ \xpom )^{\pm 0.1}$, 
which leads to an average uncertainty of 2.0\% in $\sigma_r^D$.
The $\beta$ distribution is reweighted by $\beta^{\pm 0.1}$ and 
$(1- \beta )^{\pm 0.1}$, leading to typical uncertainties of 3.2\%.
Reweighting the
$t$ distribution by $e^{\pm t}$ results in uncertainties of 2.5\% on average.

\item The uncertainties related to the subtraction of 
backgrounds (see section~\ref{mc}) are  
at most $2.7\%$ for proton dissociation,
$3.0\%$ for photoproduction and $3.5\%$ for the 
proton beam-halo contribution.
  
\item A $2.6\%$ uncertainty is 
attributed to the trigger efficiencies (section~\ref{recsec}), evaluated using 
independent triggers.

\item The uncertainty in the FPS track reconstruction efficiency 
results in an overall normalisation uncertainty of 
$10\%$ (see section~\ref{detector}). 
A further normalisation uncertainty of $1.5\%$ arises from 
the luminosity measurement.

\item The extrapolation from the measured FPS
range of $0.08 < |t|  < 0.5~\GeV^2$ to the region 
$|t_{\rm min}| < |t| < 1~{\GeV}^2$ covered by the LRG data \cite{H1LRG}
results in an additional systematic error of up to 
$5\%$ for the $\sigma_r^{D(3)}$ data (see section~\ref{f2d3sec}).

\end{itemize}

\noindent
The systematic errors shown in the figures and tables are calculated as the 
quadratic sum of all contributions which vary from point to point, 
corresponding to average uncertainties of $12\%$ for the
$\sigma_r^{D(4)}$ data and
$13\%$ for $\sigma_r^{D(3)}$.
The quoted errors do not include the overall 
normalisation uncertainty.


\section{Results and Discussion}

\subsection{Cross section dependence on {\boldmath $t$}}
\label{tdep}

\begin{figure}[p] \unitlength 1mm
 \begin{center}
 \begin{picture}(160,190)
    \put(18,95){\epsfig{file=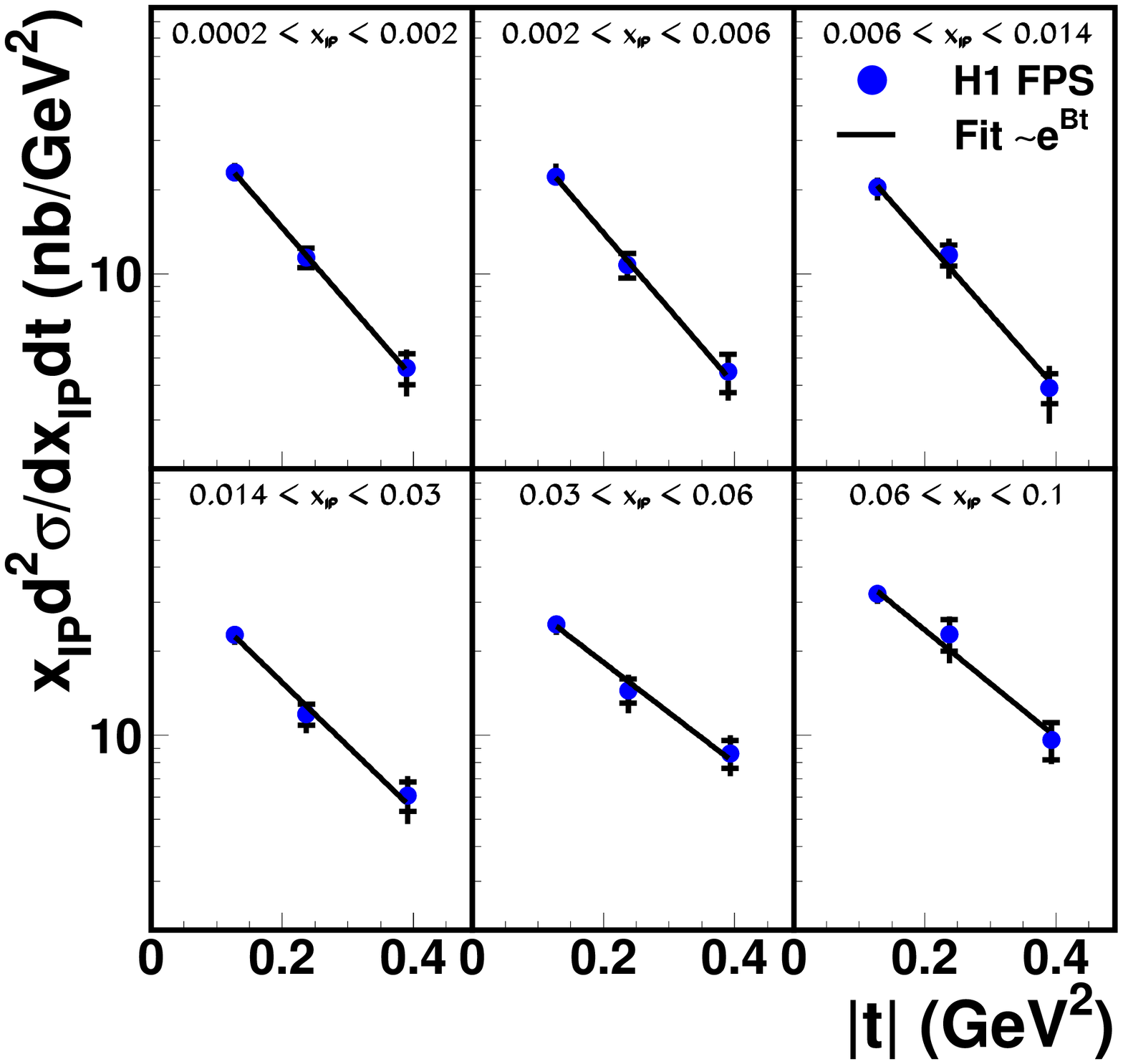,width=0.63\linewidth}}
    \put(18,-5){\epsfig{file=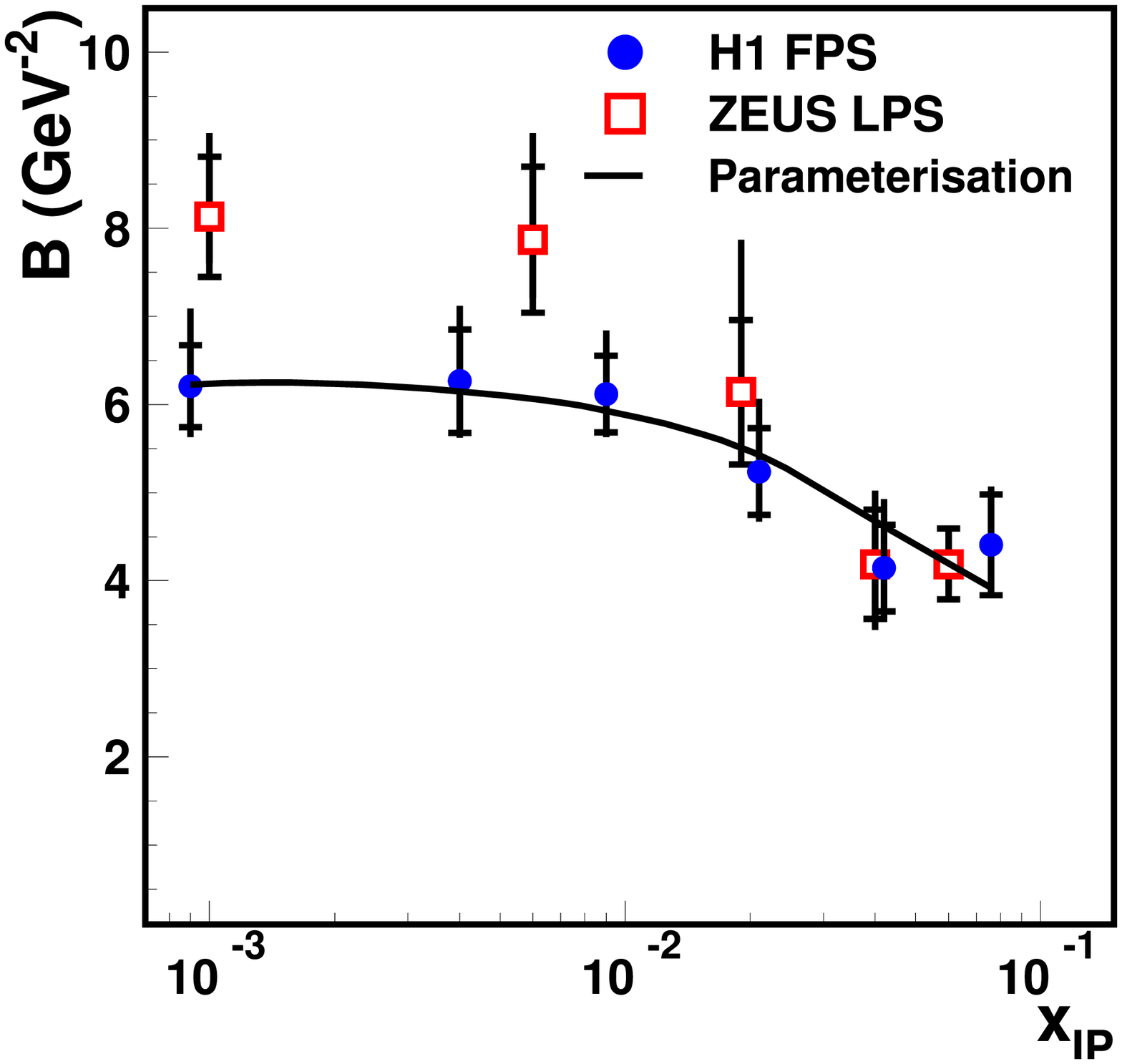,width=0.63\linewidth}}
    \put(140,160){\bf{\Large{(a)}}}
    \put(140,53){\bf{\Large{(b)}}}
 \end{picture}
 \end{center}
 \caption{(a) The  differential cross section $\xpom \, {\rm d}^2 \sigma / {\rm d}\xpom {\rm d}t$ measured in the kinematic 
range $2<Q^2<50~{\GeV}^2, 0.02<y<0.6$ for different $\xpom$ intervals.
The results of fits of the form
$\xpom {\rm d}^2 \sigma /{\rm d} \xpom {\rm d}t \propto e^{Bt}$ are 
also shown. 
(b) The slope parameter $B$  
obtained from these fits, shown as a function of $\xpom$. 
The results obtained with the ZEUS LPS~\cite{ZEUSLPS} 
and the parameterisation of the H1 data described in 
section~\ref{f2d4sec} are also shown.
The inner error bars represent the statistical errors and
the outer error bars indicate the statistical and systematic errors added in 
quadrature.}
\label{fig:tdepxp}
\end{figure}

The  differential cross section 
${\rm d}^2 \sigma / {\rm d}\xpom \, {\rm d}t$
provides a measurement of the $t$ dependence of diffractive
DIS. This cross section
is shown in figure~\ref{fig:tdepxp}a, 
multiplied by $\xpom$ for convenience, for 
three values of $t$ and six values of $\xpom$ in the range
$\xpom <0.1$ and $0.08 < |t|  < 0.5~\GeV^2$,
integrated over
$2< Q^2 <50~\GeV^2$ and $0.02< y <0.6$.
For each $\xpom$ value, fits to the form
$\xpom \ {\rm d}^2 \sigma / {\rm d}\xpom {\rm d}t \propto e^{Bt}$
are shown in figure~\ref{fig:tdepxp}a. The extracted values of the 
slope parameter $B$ are plotted
as a function of $\xpom$ in figure~\ref{fig:tdepxp}b and are listed in 
table~\ref{table:tdep}. 
The H1 results for $B$ are consistent 
with ZEUS measurements~\cite{ZEUSLPS}, though the H1 data are somewhat
lower than the ZEUS data for
$\xpom \lapprox 0.02$.

At low $\xpom$, the data are compatible with a constant
slope parameter, $B \simeq 6 \ {\rm GeV^2}$. 
In a Regge approach with a single linear exchanged
trajectory, $\alpha_{\pom}(t) = \alpha_{\pom}(0) + \alpha_\pom^\prime t$, the
slope parameter
is expected to decrease logarithmically with increasing
$\xpom$ according to 
\begin{eqnarray}
B = B_\pom - 2\alpha_{\pom}^{\prime}\ln \xpom \ ,
\label{tslope}
\end{eqnarray}
an effect which is often referred to as `shrinkage' of the diffractive 
peak. The degree of shrinkage depends on the slope of the pomeron trajectory, 
which is
$\alpha_{\pom}^{\prime} \simeq 0.25~\GeV^{-2}$ 
for soft hadron-hadron scattering at high 
energies~\cite{softpom}. In contrast, vector meson measurements at HERA
have resulted in smaller values of $\alpha_{\pom}^{\prime}$, 
whether a hard scale is present \cite{zeus:jpsi,h1:jpsi}
or not \cite{zeus:rho}.
Fits of the form of equation~\ref{tslope} are performed 
to the FPS data shown in figure~\ref{fig:tdepxp}b
in the region where pomeron exchange is expected to dominate, namely to
the three data points with $0.0009 \leq \xpom \leq 0.0094$, 
for which the sub-leading exchange 
contribution is estimated to be at most $7\%$ (see the fit
results in section~\ref{f2d4sec}). 
A two parameter 
fit to the data in this range 
yields 
$B_\pom=6.0 \pm 1.6 \ {\rm (stat.)} _{-1.0}^{+2.4} \ {\rm (syst.)} \ \GeV^{-2}$ and 
$\alpha_\pom^{\prime} = 0.02 \pm 0.14 \ {\rm (stat.)} _{-0.09}^{+0.21} \ {\rm (syst.)} \ \GeV^{-2}$.
Extending the fit range to the interval $0.0009 \leq \xpom \leq 0.021$,
for which the contribution of the sub-leading exchange is at most
$20\%$ (section~\ref{f2d4sec}), results in 
$B_\pom=4.9 \pm 1.2 \ {\rm (stat.)} _{-0.7}^{+1.6} \ {\rm (syst.)} \ \GeV^{-2}$ 
and $\alpha_\pom^{\prime} = 0.10 \pm 0.10 \ {\rm (stat.)} _{-0.07}^{+0.16} \ 
{\rm (syst.)} \ \GeV^{-2}$. 
The data thus favour a small value of $\alpha^\prime$,
as expected in perturbative models of the 
pomeron \cite{bfkl}.
However, the result $\alpha_\pom^{\prime} \simeq 0.25$ from soft interactions 
cannot be excluded. The results of these fits are summarised in 
table~\ref{table:alphaprime}.
   
\renewcommand{\arraystretch}{1.35}
\begin{table}[h]
\centering
\begin{tabular}{|l|l|l|}
\hline
\multicolumn{1}{|c|}{Range of Fit} & 
\multicolumn{1}{c|}{$\alpha_\pom^\prime$ ($\rm GeV^{-2}$)} & 
\multicolumn{1}{c|}{$B_\pom$ ($\rm GeV^{-2}$)} \\
\hline
$0.0009 \leq \xpom \leq 0.0094$ & $0.02 \pm 0.014 ^{+0.21}_{-0.09}$ 
                              & $6.0  \pm 1.6   ^{+2.4}_{-1.0}$ \\
$0.0009 \leq \xpom \leq 0.021$ & $0.10 \pm 0.010 ^{+0.16}_{-0.07}$ 
                              & $4.9  \pm 1.2   ^{+1.6}_{-0.7}$ \\
\hline
\end{tabular}
\caption{The results of fits to the slope parameter data in two
different ranges at low $\xpom$ in order to extract $\alpha_\pom'$ and
$B_\pom$, together with their statistical (first error) and systematic
(second error) uncertainties.}
\label{table:alphaprime}
\end{table}
\renewcommand{\arraystretch}{1.}

A decrease of the slope $B$ is observed 
towards the region of larger $\xpom \gapprox 0.03$, 
where the contribution
from the sub-leading exchange is expected to be 
significant ($60\%$ in the highest bin at $\xpom = 0.076$). 
This reduction of the slope parameter indicates that the 
size of the interaction region reduces as $\xpom$ increases,
reaching values of around $4 \ {\rm GeV^{-2}}$, characteristic
of the spatial extent of the proton charge distribution. 

\begin{figure}[p] \unitlength 1mm
 \begin{center}
 \begin{picture}(160,190)
    \put(-3,95){\epsfig{file=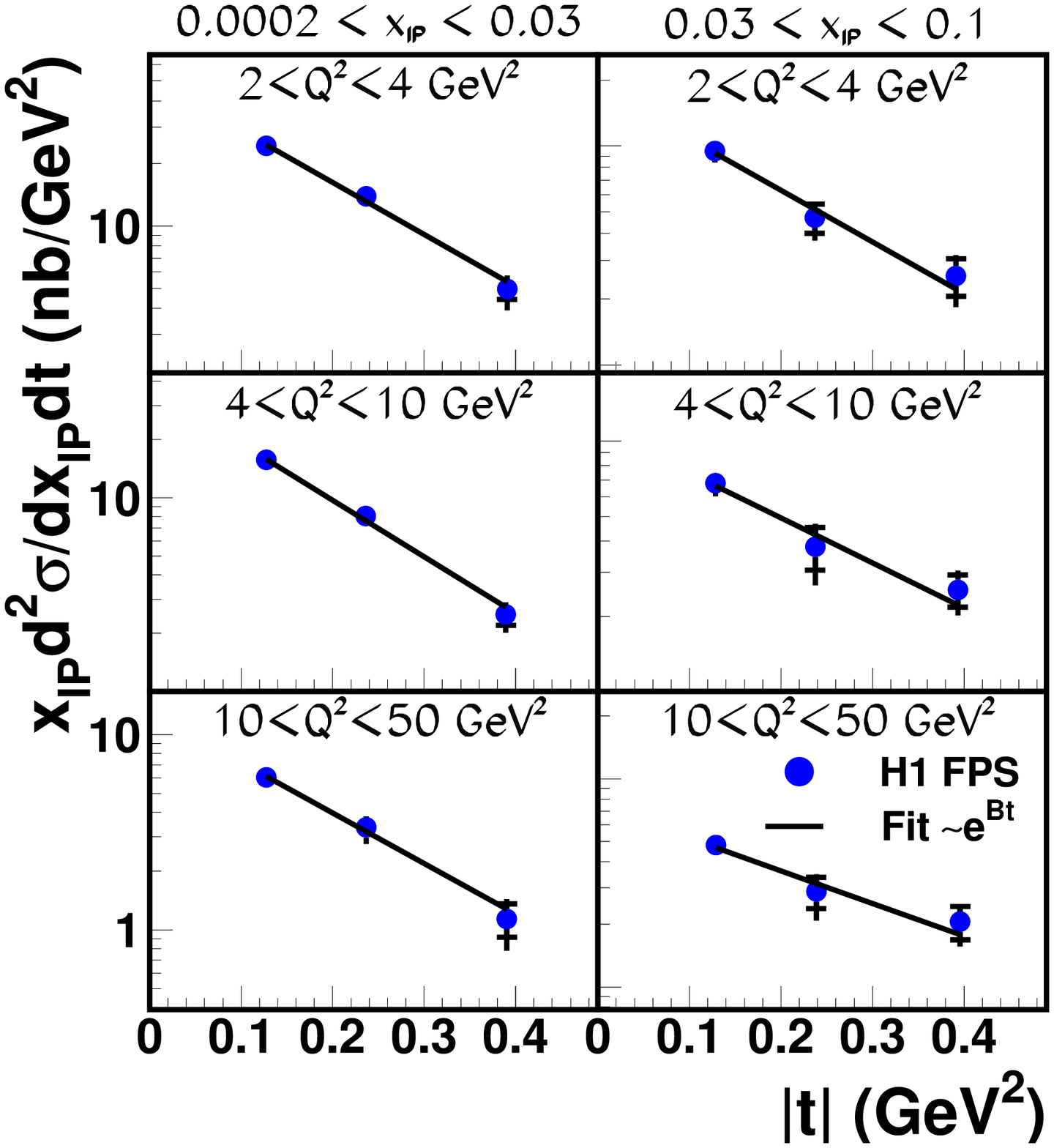,width=0.5\linewidth}}
    \put(80,95){\epsfig{file=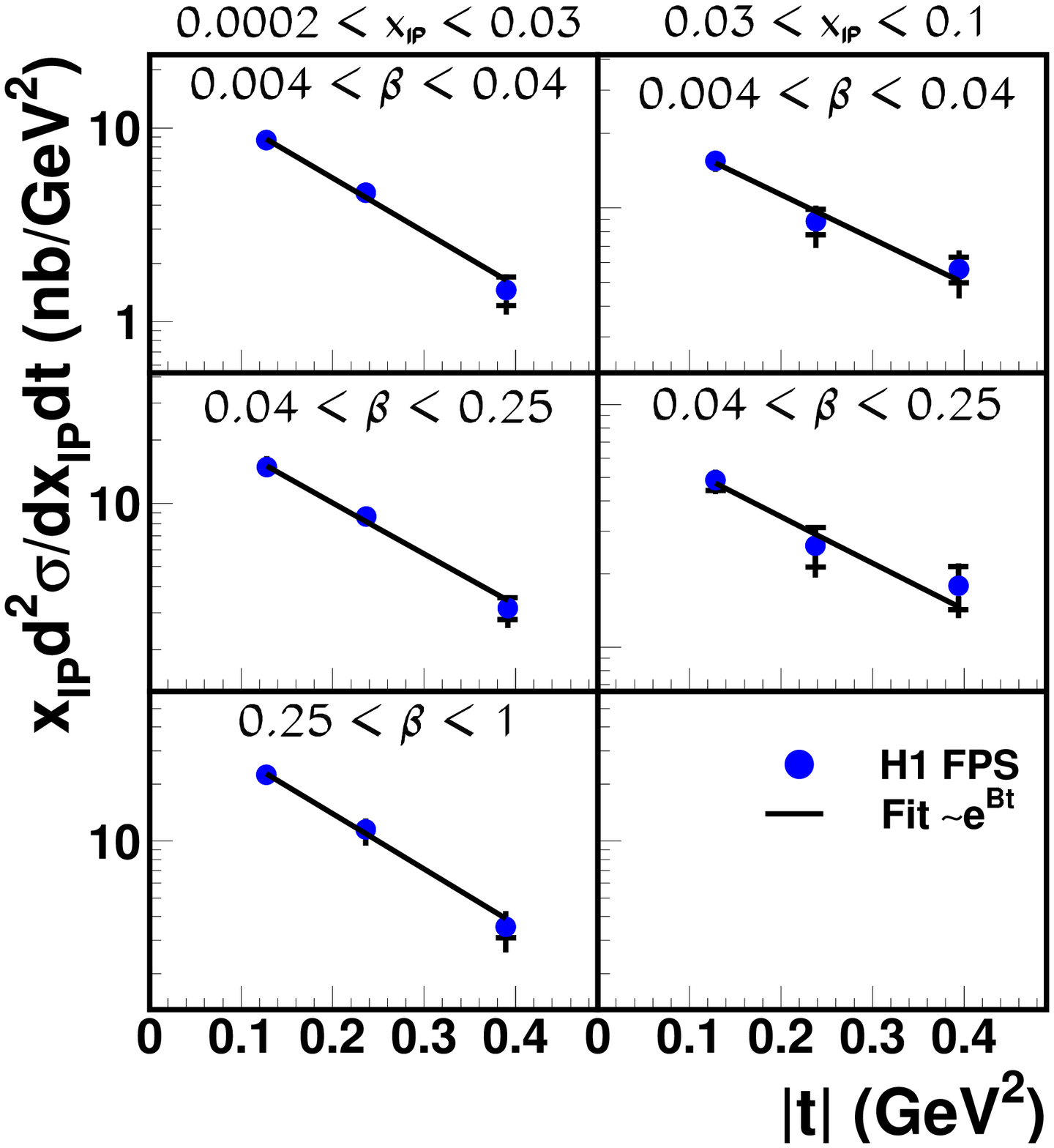,width=0.5\linewidth}}
    \put(-3,-3){\epsfig{file=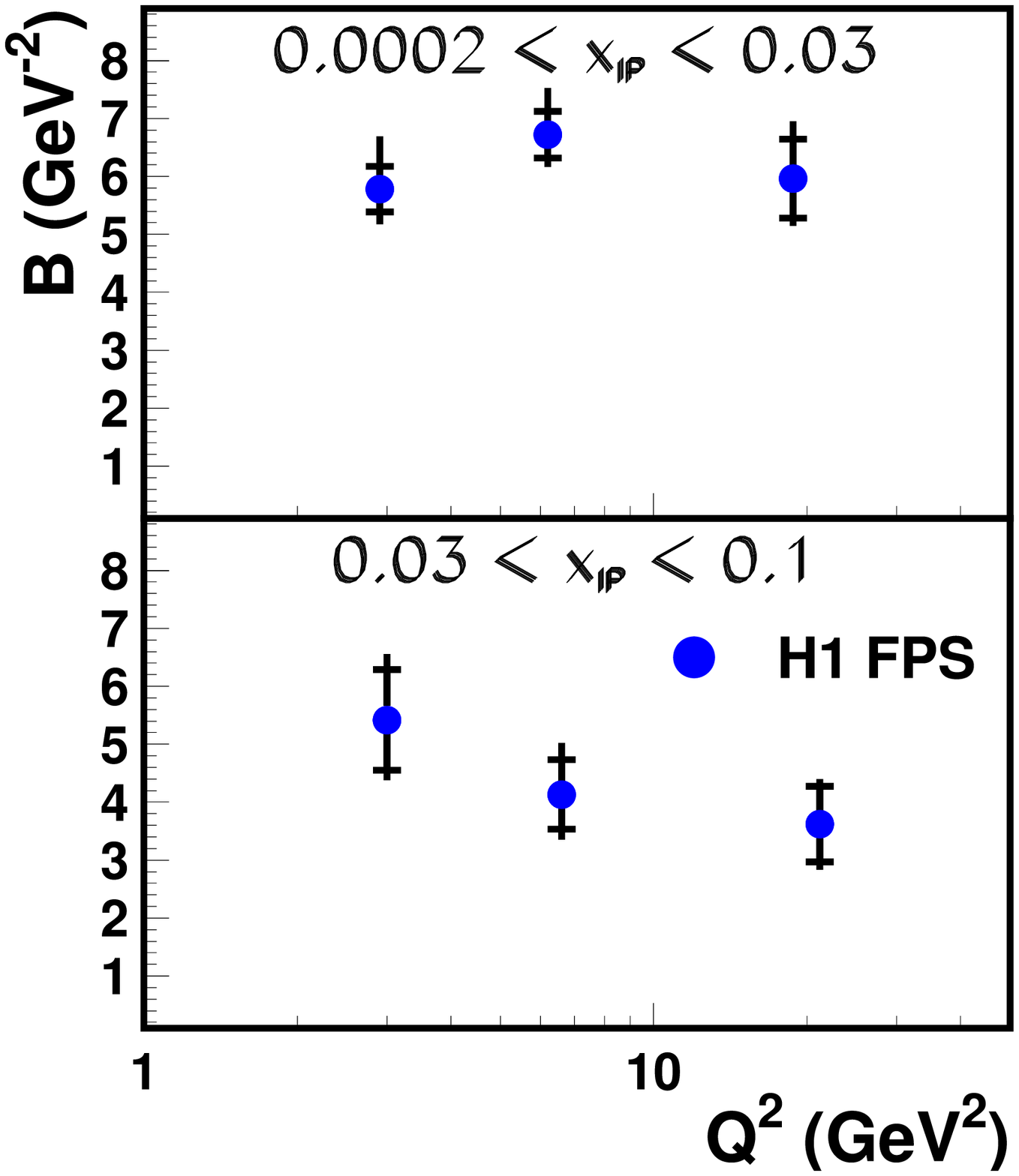,width=0.5\linewidth}}
    \put(80,-3){\epsfig{file=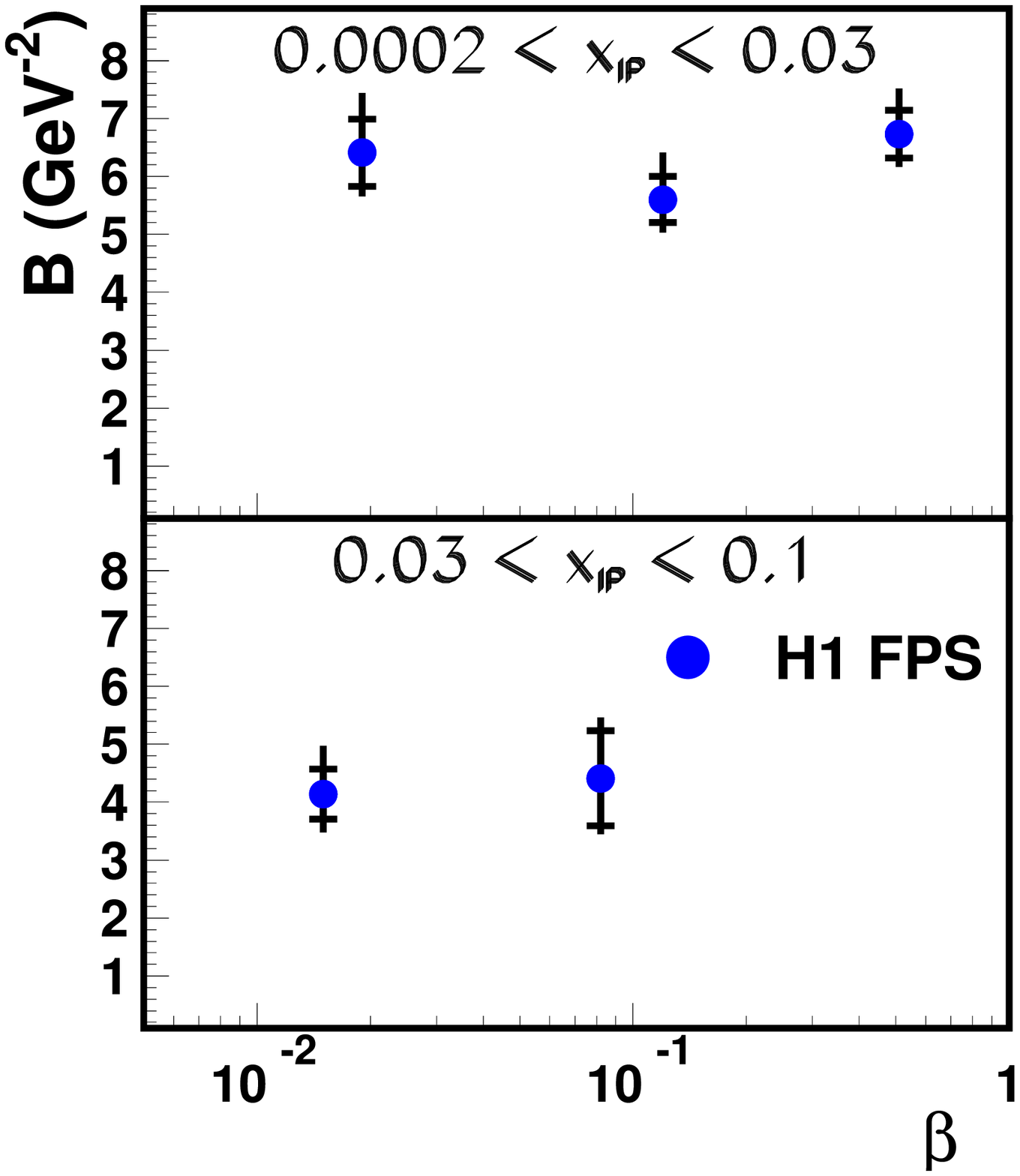,width=0.5\linewidth}}
    \put(40,185){\bf{\Large{(a)}}}
    \put(120,185){\bf{\Large{(b)}}}
    \put(40,88){\bf{\Large{(c)}}}
    \put(120,88){\bf{\Large{(d)}}}
 \end{picture}
 \end{center}
 \caption{(a,b) The  differential cross section 
$\xpom \, {\rm d}^2 \sigma / {\rm d}\xpom {\rm d}t$ measured in different 
regions of (a) $Q^2$ and $\xpom$ and 
(b) $\beta$ and $\xpom$.
The results of fits of the form
$\xpom {\rm d}^2 \sigma /{\rm d} \xpom {\rm d}t \propto e^{Bt}$ are shown. 
(c,d) The slope parameter $B$ obtained from these fits, shown 
as a function of (c) $Q^2$ and (d) $\beta$ for two $\xpom$ intervals.
The inner error bars represent the statistical errors and
the outer error bars indicate the statistical and systematic errors added in 
quadrature.}
\label{fig:tdepbq2}
\end{figure}

The $t$ dependence of the cross section is 
also presented in figure~\ref{fig:tdepbq2} and 
table~\ref{table:tdep2}
in different regions of $Q^2$ and $\beta$ for two $\xpom$ intervals. 
No significant $Q^2$ or $\beta$
dependence of the slope parameter $B$ is observed for $\xpom < 0.03$.
Within the uncertainties, 
the $t$ dependence of the cross section in the pomeron dominated low $\xpom$
region
can therefore be factorised from the 
$Q^2$ and $\beta$ dependences.
Since there is also no strong evidence for any $\beta$ or $Q^2$ dependence
of $B$ for $\xpom > 0.03$, the data are consistent with a similar
factorisation for the sub-leading 
exchange contribution.

\subsection{Cross section dependence on {\boldmath $\xpom$} and
extraction of {\boldmath $\alpha_\pom(0)$}}
\label{f2d4sec}

The $\xpom$, $\beta$ and $Q^2$ dependences of diffractive DIS 
are studied in terms of the diffractive reduced cross sections
$\sigma_r^{D(4)}$ and $\sigma_r^{D(3)}$.  
The former observable is related to the measured differential
cross section by \cite{H1LRG} 
\begin{eqnarray}
 \frac{{\rm d}^4 \sigma^{ep \rightarrow eXp}}{{\rm d}\beta dQ^2 {\rm d}\xpom {\rm d}t} = \frac{4 \pi \alpha^2}{\beta Q^4} \cdot 
  \left( 1 - y + \frac{y^2}{2} \right)
  \cdot \sigma_r^{D(4)}(\beta,Q^2,\xpom,t) \ .
 \label{sigrddef}
\end{eqnarray}
The reduced cross section is equal to the diffractive structure function
$F_2^{D(4)} (\beta,Q^2,\xpom,t)$ to good approximation in the 
relatively low $y$ region
covered by the current analysis, where the contribution from the
longitudinal structure function
$F_L^{D(4)} (\beta,Q^2,\xpom,t)$ is small.
Results for $\sigma_r^{D(4)}$ are obtained at a fixed
value of $|t|=0.25 \ {\rm \GeV^2}$, interpolating from the measured
range $0.08 < |t| < 0.5~\GeV^2$
using the measured $t$ dependence at each $\xpom$
value (figure~\ref{fig:tdepxp}).
Presenting the measurement at $|t|=0.25 \ {\rm \GeV^2}$ ensures that
the systematic uncertainties associated with this interpolation
are small. 

\begin{figure}[p] \unitlength 1mm
 \begin{center}
 \begin{picture}(160,180)
    \put(-3,0){\epsfig{file=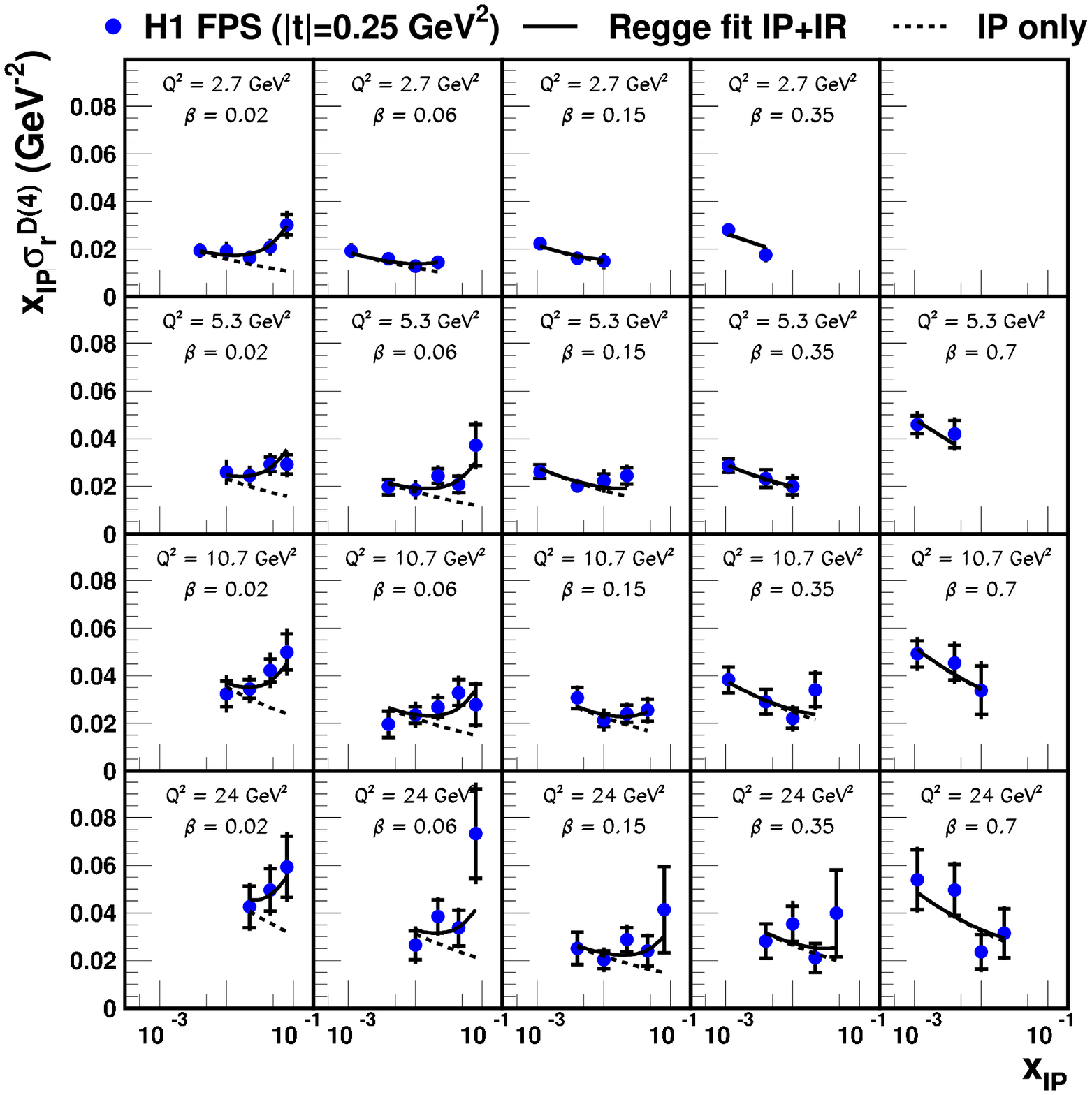,width=\linewidth}}
 \end{picture}
 \end{center}
 \caption{The  diffractive 
reduced cross section 
$\xpom \, \sigma_r^{D(4)}(\beta,Q^2,\xpom,t)$, shown
as a function of $\xpom$ for 
$|t|=0.25~\GeV^2$
at different values of $\beta$ and $Q^2$.  
The inner error bars represent the statistical errors.
The outer error bars indicate the statistical and systematic errors 
added in quadrature. An overall
normalisation uncertainty of 10.1\% is not shown.
The solid curves represent the results of the phenomenological 
`Regge' fit to 
the data, including both pomeron ($\pom$) and sub-leading ($\reg$)
trajectory exchange,  
as described in section~\ref{f2d4sec}. 
The dashed curves represent the contribution
from pomeron exchange alone according to the fit.}
\label{fig:f2d4}
\end{figure}

Figure~\ref{fig:f2d4} shows $\xpom \, \sigma_r^{D(4)}$
for $|t|=0.25~\GeV^2$
as a function of $\xpom$ for different $Q^2$ and $\beta$ values 
(see also tables~\ref{table:f2d1}-~\ref{table:f2d4}). 
At medium and large $\beta$ values, $\xpom \, \sigma_r^{D(4)}$ 
falls or is flat as a function
of $\xpom$. Qualitatively this behaviour is consistent with a dominant pomeron contribution with an intercept
$\alpha_{\pom}(0) \gtrsim 1$.
However, $\xpom \sigma_r^{D(4)}$ rises with $\xpom$ at 
the highest $\xpom$ for low $\beta$ values, which can be 
explained by a contribution from a sub-leading exchange
with an intercept $\alpha_\reg(0) < 1$.

To describe the $\xpom$ dependence quantitatively,
a fit is performed to the structure function $F_2^{D(4)}$, obtained 
by correcting $\sigma_r^{D(4)}$
for the small $F_L^{D(4)}$ contribution using the results
of the `2006 DPDF fit A' in \cite{H1LRG}. 
A parameterisation of the form
\begin{eqnarray}
 F_2^{D(4)} = f_{\pom}(\xpom,t) F_{\pom}(\beta,Q^2) + n_{\reg} \cdot f_{\reg}(\xpom,t) F_{\reg}(\beta,Q^2) 
\label{regfit}
\end{eqnarray}
is used.
This parameterisation assumes a separate `proton vertex' factorisation 
of the $\xpom$ and $t$ dependences
from those on $\beta$ and $Q^2$ for both the 
pomeron and a sub-leading exchange.
The factors 
$f_{\pom}$ and $f_{\reg}$ correspond to  
flux factors for the exchanges and are taken from the Regge-motivated 
functions,
\begin{eqnarray}
 f_{\pom}(\xpom,t) = A_{\pom} \, . \, \frac{e^{B_{\pom} \, t}}{\xpom^{2\alpha_{\pom}(t)-1}} \ \ \  \ \ ; \ \ \ \ \ 
 f_{\reg}(\xpom,t) = A_{\reg} \, . \, \frac{e^{B_{\reg} \, t}}{\xpom^{2\alpha_{\reg}(t)-1}} \ ,
\label{fluxfac}
\end{eqnarray}
assuming that the sub-leading exchange
has a linear trajectory
$\alpha_\reg(t) = \alpha_\reg(0) + \alpha_\reg^\prime t$ as for the
pomeron.
The values of $A_{\pom}$ and $A_{\reg}$ are chosen such that 
$\xpom \cdot \int_{-1}^{t_{\rm min}} f_{\pom,\reg}(\xpom,t) \, {\rm d} t = 1$ 
at $\xpom = 0.003$, following 
the convention of \cite{H1Diff94}.
The free parameters of the fit are 
the pomeron intercept $\alpha_{\pom}(0)$,
normalisation coefficients $F_{\pom}(\beta,Q^2)$ for the pomeron 
contribution at each of the nineteen ($\beta, Q^2$) values considered,
and a single parameter $n_{\reg}$ describing the
normalisation of the sub-leading exchange contribution.

A summary of the values assumed for the parameters which are fixed in the
fits is given in table~\ref{table:params}. 
The intercept $\alpha_\reg (0)$ of the sub-leading exchange is 
obtained from \cite{H1Diff94}. 
As in \cite{H1Diff94,H1LRG}, 
the normalisation 
coefficients $F_{\reg}(\beta,Q^2)$ for the sub-leading exchange 
in each $\beta$ and $Q^2$ bin are taken from a 
parameterisation of the pion structure 
function~\cite{Owens}.
The remaining fixed parameters describing the fluxes
are taken from the present analysis. Averages of the two fits 
to the
$B(\xpom)$ data at low $\xpom$ 
described in section~\ref{tdep} (table~\ref{table:alphaprime}) 
are used 
to fix the pomeron parameters, $B_\pom = 5.5 \ {\rm GeV^{-2}}$ and 
$\alpha_\pom^\prime = 0.06 \ {\rm GeV^{-2}}$.
The behaviour of $B(\xpom)$ 
at large $\xpom$ is sensitive to the parameters
$\alpha_\reg^\prime$ and $B_\reg$. Although the constraints are not
strong, the data are incompatible with the pair of values, 
$\alpha_\reg^\prime = 0.9 \ {\rm GeV^{-2}}$ 
\cite{apel:chexch} and
$B_\reg = 2.0 \ {\rm GeV^{-2}}$~\cite{kaidalov}, 
obtained from soft hadronic scattering
data and applied previously in similar fits to $F_2^D$ 
data~\cite{H1Diff94}. 
A good description of the slope parameter results over the full $\xpom$
range is obtained with
$B_\reg = 1.6 \ {\rm GeV^{-2}}$ and 
$\alpha_\reg^\prime = 0.3 \ {\rm GeV^{-2}}$. A description based on
this parameterisation
is shown in figure~\ref{fig:tdepxp}b. 

\renewcommand{\arraystretch}{1.35}
\begin{table}[h]
\centering
\begin{tabular}{|l|r@{$\,$}l|}
\hline
Parameter & \multicolumn{2}{c|}{Value} \\
\hline
$\alpha_\pom'$    & $0.06$ & $^{\,+\,0.19}_{\,-\,0.06} \rm\ GeV^{-2}$ \\
$B_\pom$          & $5.5$ & $^{\,-\,2.0}_{\,+\,0.7} \rm\ GeV^{-2}$    \\
$\alpha_\reg(0)$  & $0.50$ & $ \pm \,0.10$                            \\
$\alpha_\reg'$    & $0.3$ & $^{\,+\,0.6}_{\,-\,0.3} \rm\ GeV^{-2}$    \\
$B_\reg$          & $1.6$ & $^{\,-\,1.6}_{\,+\,0.4} \rm\ GeV^{-2}$    \\
\hline
\end{tabular}
\caption{The values of the fixed parameters
and their uncertainties, as used in the extraction of $\alpha_\pom (0)$. 
Since $\alpha_\pom'$ and
$B_\pom$ are strongly
anti-correlated when extracted
from the data shown in figure~\ref{fig:tdepxp}b, 
they are varied simultaneously to obtain the
errors on the fit
results, as are $\alpha_\reg'$ and
$B_\reg$.}
\label{table:params}
\end{table}
\renewcommand{\arraystretch}{1.}

The experimental systematic uncertainties on the free parameters 
are evaluated by repeating the fit after shifting the data
points according to each individual uncertainty source
described in section~\ref{systs}.
A model dependence uncertainty is determined by varying the 
fixed parameters as described in table~\ref{table:params}.
The $\alpha^\prime_{\pom,\reg}$ and
$B_{\pom,\reg}$ parameters
are varied
in the ranges given in the table, within which an acceptable
description of the data is maintained, whilst requiring that 
$\alpha_\pom^\prime$ and $\alpha_\reg^\prime$ lie between $0$ and the
values describing soft hadronic scattering ($0.25 \ {\rm GeV^{-2}}$ and 
$0.9 \ {\rm GeV^{-2}}$, respectively). 
The influence of neglecting the $F_L^{D(4)}$ contribution 
is also included in the model dependence uncertainty.
 
As shown in figure~\ref{fig:f2d4},
the fit provides
a good description of the $\xpom$ dependence of the data
($\chi^2 = 44$ with statistical uncertainties 
for 51 degrees of freedom).
Within uncertainties, the $\xpom$ dependence
can therefore be factorised from the $\beta$ and $Q^2$ 
dependences for each of the pomeron and the sub-leading contributions. 

The fit yields a pomeron intercept of 
\begin{eqnarray*}
\alpha_{\pom}(0)=1.114 \pm 0.018 \ ({\rm stat.}) 
\ \pm \ 0.012 \ ({\rm syst.}) \
^{+0.040}_{-0.020} \ ({\rm model}) \ , 
\end{eqnarray*}
the dominant uncertainty arising
from the variations of $\alpha_\pom^\prime$ and $B_\pom$.
This result for $\alpha_\pom (0)$ is 
compatible with that obtained from H1 
data measured using the LRG
method~\cite{H1LRG} and with 
ZEUS measurements~\cite{ZEUSLPS,ZEUSMX}. 
It is 
only slightly higher than the pomeron intercept 
describing soft hadronic scattering, 
$\alpha_{\pom}(0) \simeq 1.08$ \cite{softpom}. However, if $\alpha_\pom^\prime$
is set to the soft pomeron value of $0.25 \ {\rm GeV^{-2}}$, $\alpha_\pom(0)$
increases to around 1.15.

The result for the sub-leading 
exchange normalisation parameter is
\begin{eqnarray*} 
n_{\reg} \ = \ [1.0 \ \pm \ 0.2 \ {\rm (stat.)} \ \pm \ 0.1 \ {\rm (syst.)} \
^{+1.2}_{-0.7} \ {\rm (model)} \, ] \times 10^{-3} \ ,
\end{eqnarray*} 
the largest uncertainty arising from the
variation of $\alpha_\reg(0)$.
The sub-leading exchange
is important at low $\beta$ and high $\xpom$,
contributing typically $60\%$ of the cross section
at the highest $\xpom = 0.08$. 

\subsection{Cross section dependence on 
{\boldmath $Q^2$} and {\boldmath $\beta$}}
\label{f2d3sec}

\begin{figure}[p] \unitlength 1mm
 \begin{center}
 \begin{picture}(160,180)
    \put(-3,0){\epsfig{file=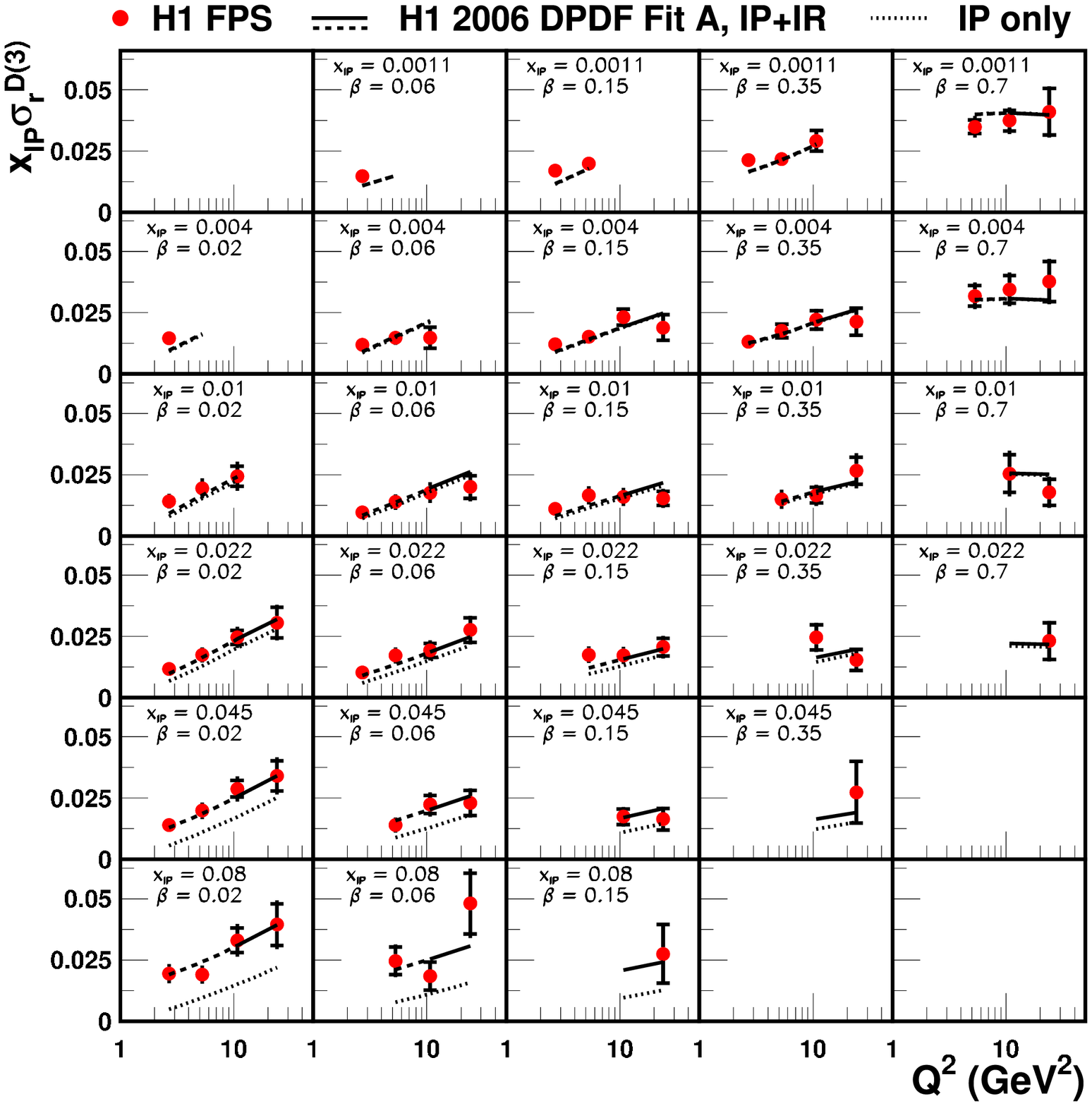,width=\linewidth}}
 \end{picture}
 \end{center}
 \caption{The  diffractive reduced cross section
$\xpom \, \sigma_r^{D(3)}(\beta,Q^2,\xpom)$
for $|t| < 1 \ {\rm GeV^2}$,
shown as a function of $Q^2$ for
different values of $\xpom$ and $\beta$.
The inner error bars represent the statistical errors.
The outer error bars indicate the statistical and systematic 
errors added in quadrature. An overall
normalisation uncertainty of 10.1\% is not shown.
The solid curves represent the results of the `H1 2006 DPDF  
Fit A' to LRG data \cite{H1LRG}, modified as
described in 
section~\ref{lrgcomp}. 
The dashed curves represent the extrapolation 
of this prediction beyond 
the $Q^2$ range which is included in the fit.
The dotted curves indicate the contribution 
of pomeron exchange alone in this model.}
\label{fig:f2d3q2} 
\end{figure}

\begin{figure}[p] \unitlength 1mm
 \begin{center}
 \begin{picture}(160,180)
    \put(-3,0){\epsfig{file=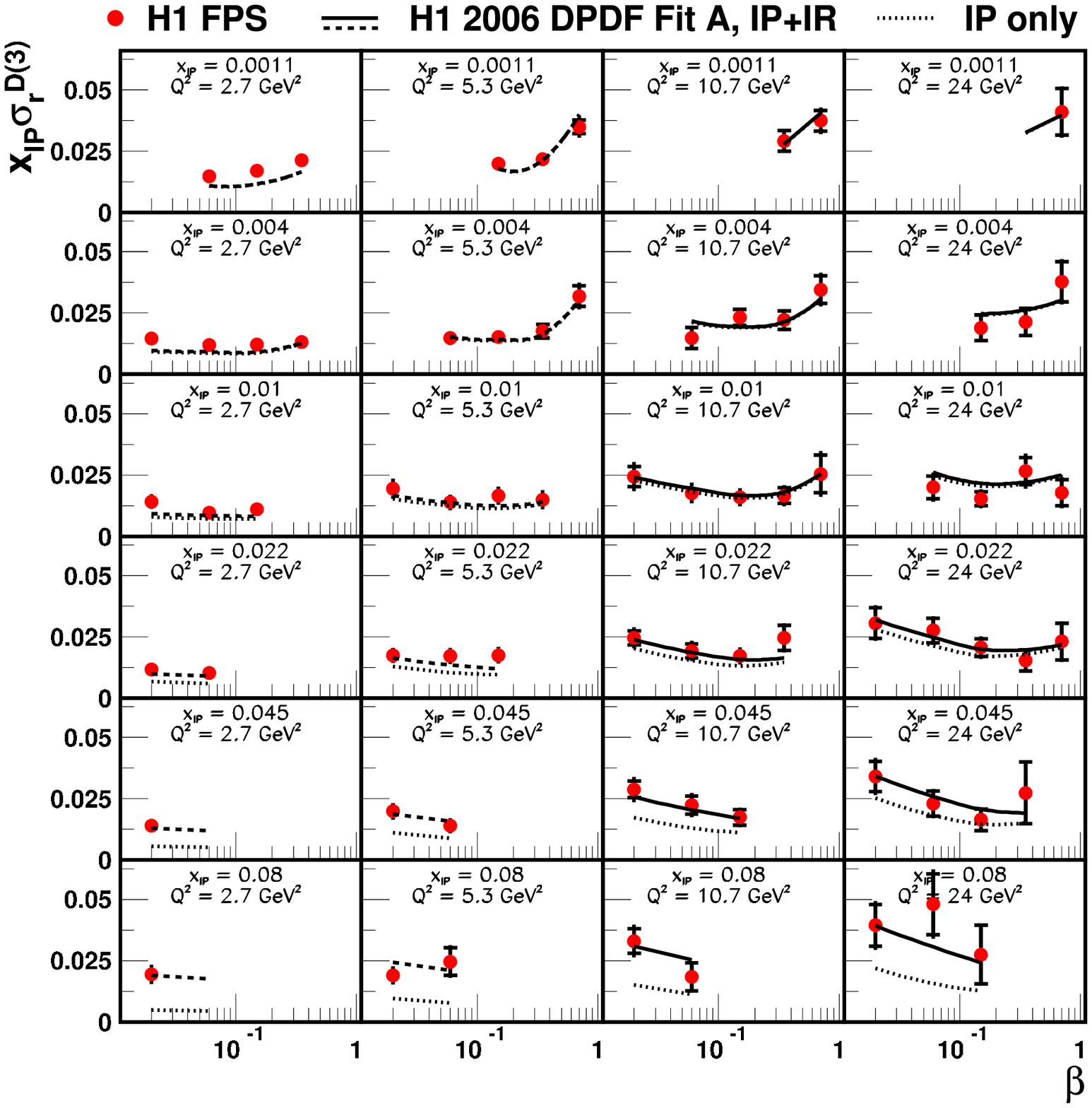,width=\linewidth}}
 \end{picture}
 \end{center}
 \caption{The  diffractive reduced cross section
$\xpom \, \sigma_r^{D(3)}(\beta,Q^2,\xpom)$
for $|t| < 1 \ {\rm GeV^2}$,
shown as a function of $\beta$ for
different values of $\xpom$ and $Q^2$.
The inner error bars represent the statistical errors.
The outer error bars indicate the statistical and systematic 
errors added in quadrature. An overall
normalisation uncertainty of 10.1\% is not shown.
The solid curves represent the results of the `H1 2006 DPDF  
Fit A' to LRG data \cite{H1LRG}, modified as
described in 
section~\ref{lrgcomp}. 
The dashed curves represent the extrapolation 
of this prediction beyond 
the $Q^2$ range which is included in the fit.
The dotted curves indicate the contribution 
of pomeron exchange alone in this model.}
\label{fig:f2d3beta} 
\end{figure}

The reduced cross section $\sigma_r^{D(3)} (\beta,Q^2,\xpom)$ 
is defined as the integral of 
$\sigma_r^{D(4)} (\beta,Q^2,\xpom, t)$ over the range 
$|t_{\rm min}| < |t| < 1 \ {\rm GeV^2}$,
which is the region covered by H1 using the LRG 
method~\cite{H1LRG}.
It is obtained here by extrapolating the FPS data from the measured
range $0.08 < |t| < 0.5~\GeV^2$ 
using the $t$ dependence 
at each $\xpom$ value (section~\ref{tdep} and table~\ref{table:tdep}). 
The extrapolation factor depends only weakly 
on $\xpom$ and is 1.7 on average, 
with an uncertainty of up to 5\%.  The measurement of
$\xpom \sigma_r^{D(3)}$ is
presented in figures~\ref{fig:f2d3q2}-\ref{fig:f2d3xp} and 
tables~\ref{table:f2d1}-~\ref{table:f2d4}.
The data are compared with predictions derived from the 
`2006 DPDF Fit A' to the 
LRG data presented in \cite{H1LRG}, with modifications as described in 
section~\ref{lrgcomp}.

The $Q^2$ dependence of $\sigma_r^{D(3)}$ at fixed $\xpom$ and 
$\beta$ (figure~\ref{fig:f2d3q2}) is characterised by 
positive scaling 
violations ($\partial \, \sigma_r^{D(3)} / \partial \, {\rm ln}~Q^2>0$) 
throughout the kinematic range, except possibly at 
the highest $\beta = 0.7$. 
This observation is consistent with that from H1 
measurements using the LRG 
method~\cite{H1Diff94,H1LRG} and
implies a large gluonic component to the DPDFs.
As can be seen from the model comparison, 
the positive scaling violations 
may be attributed to the pomeron contribution
even at the highest $\xpom$ values, where the sub-leading
exchange is dominant.

The dependence of $\sigma_r^{D(3)}$ on $\beta$ 
is weak over most of the kinematic 
range (figure~\ref{fig:f2d3beta}).
Since the $\beta$ 
dependence is determined in the quark-parton model by 
the diffractive quark densities, this implies that the
quark densities do not decrease at the highest values of $\beta$ studied.
Indeed, $\sigma_r^{D(3)}$ clearly rises as
$\beta \rightarrow 1$ at low $Q^2$ and $\xpom$.
Within the framework of DPDFs, this can be explained in
terms of diffractive quark densities peaking at high fractional 
momenta at low $Q^2$ \cite{H1LRG,H1Diff94}.
The $\beta$ dependence of diffractive DIS has also been interpreted
in terms of the elastic scattering from the proton
of colour dipoles produced by partonic fluctuations of the virtual 
photon \cite{BEKW,BGK,Saturation}. 
In such models, the cross section at low and intermediate $\beta$ values
is dominated by 
$q \bar{q} g$ and $q \bar{q}$ fluctuations of transversely
polarised photons, respectively. The rise of 
$\sigma_r^{D(3)}$ as $\beta \rightarrow 1$ at low $Q^2$
has been interpreted in terms of $q \bar{q}$ fluctuations 
of longitudinally polarised photons \cite{Hebecker}, which are suppressed
as $Q^2$ increases. 

\section{Comparison with Other Measurements}

\subsection{Comparison with H1 large rapidity gap data.}
\label{lrgcomp}

The FPS $\sigma_r^{D(3)}$ data can be compared with H1 measurements
obtained using the LRG technique~\cite{H1LRG},
after taking into account the slightly different cross section
definitions in the two cases. Firstly, 
the cross section $ep \rightarrow eXY$ measured with the
LRG data is defined to 
include proton dissociation to any system $Y$ with a mass in
the range $M_Y<1.6~\GeV$, whereas $Y$ is defined to be a proton in the
cross section measured with the FPS.
Secondly, if there are significant
isospin-1 contributions to the sub-leading trajectory, 
charge-exchange reactions 
producing leading neutrons are expected in the LRG 
measurement, which are not present in the proton-tagged FPS data.

A point-by point comparison between the 
$\sigma_r^{D(3)}$ data obtained with the LRG and FPS
methods can be found in \cite{H1LRG}. Here, the level of agreement is
scrutinised in more detail in the range 
$\xpom \lapprox 0.05$, to which the LRG method is applicable.
To make the comparison
with a minimum of systematic uncertainty and to test for
differences between the
kinematic dependences of the two cross sections, 
the LRG measurement 
is repeated with an identical $Q^2$, $\beta$ and $\xpom$ 
binning to that used for the FPS data.
The ratio of the two measurements is formed for each
($Q^2$, $\beta$, $\xpom$) point and the dependences
of this ratio on each kinematic variable individually is studied by taking
statistically weighted averages over the other two variables.
Since the two data sets are 
statistically independent and
the dominant sources of systematic error are very different, 
correlations between 
the uncertainties on the FPS and LRG data are neglected.

\begin{figure}[p] \unitlength 1mm
 \begin{center}
 \begin{picture}(160,190)
    \put(-3,95){\epsfig{file=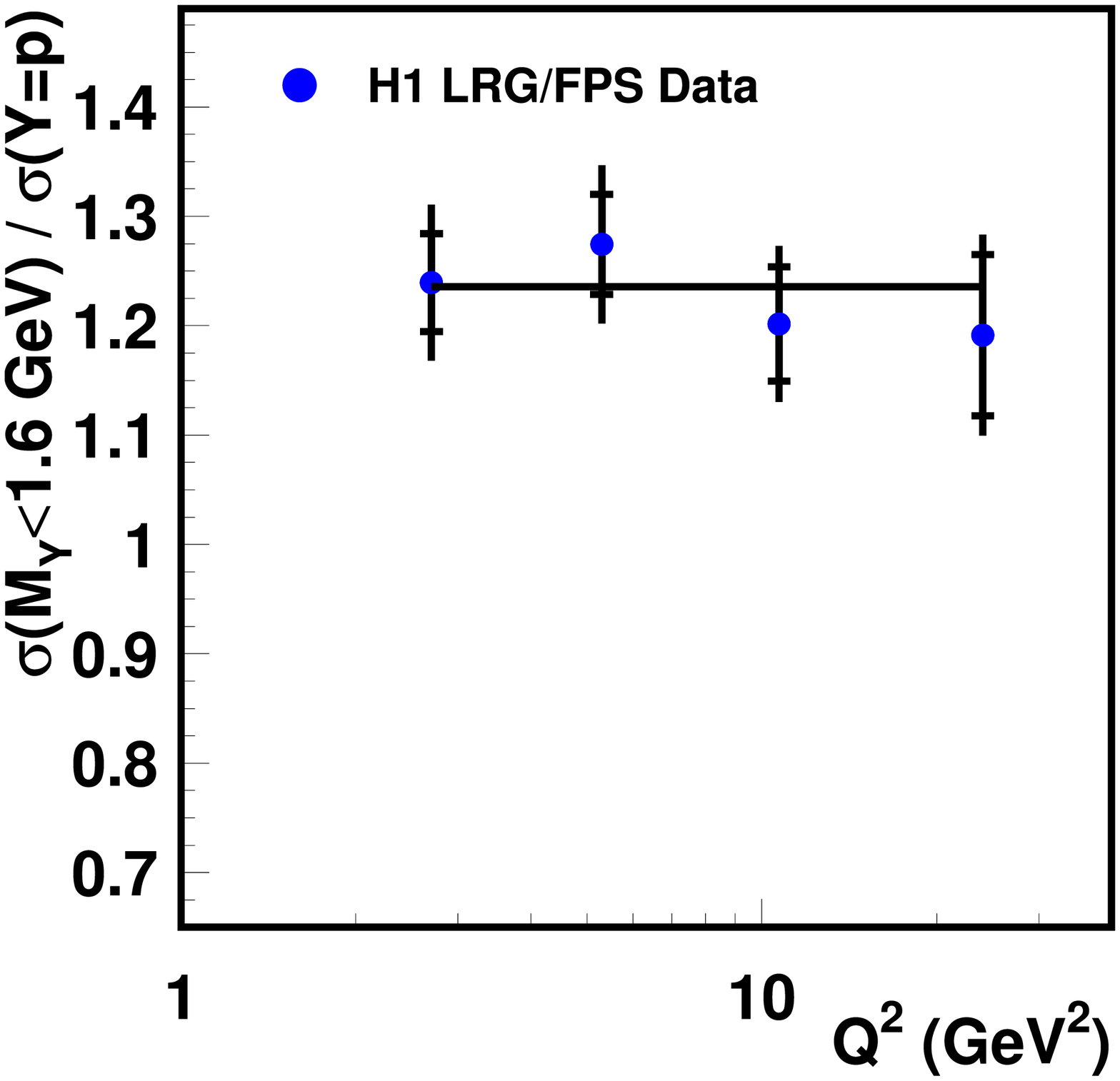,width=0.5\linewidth}}
    \put(80,95){\epsfig{file=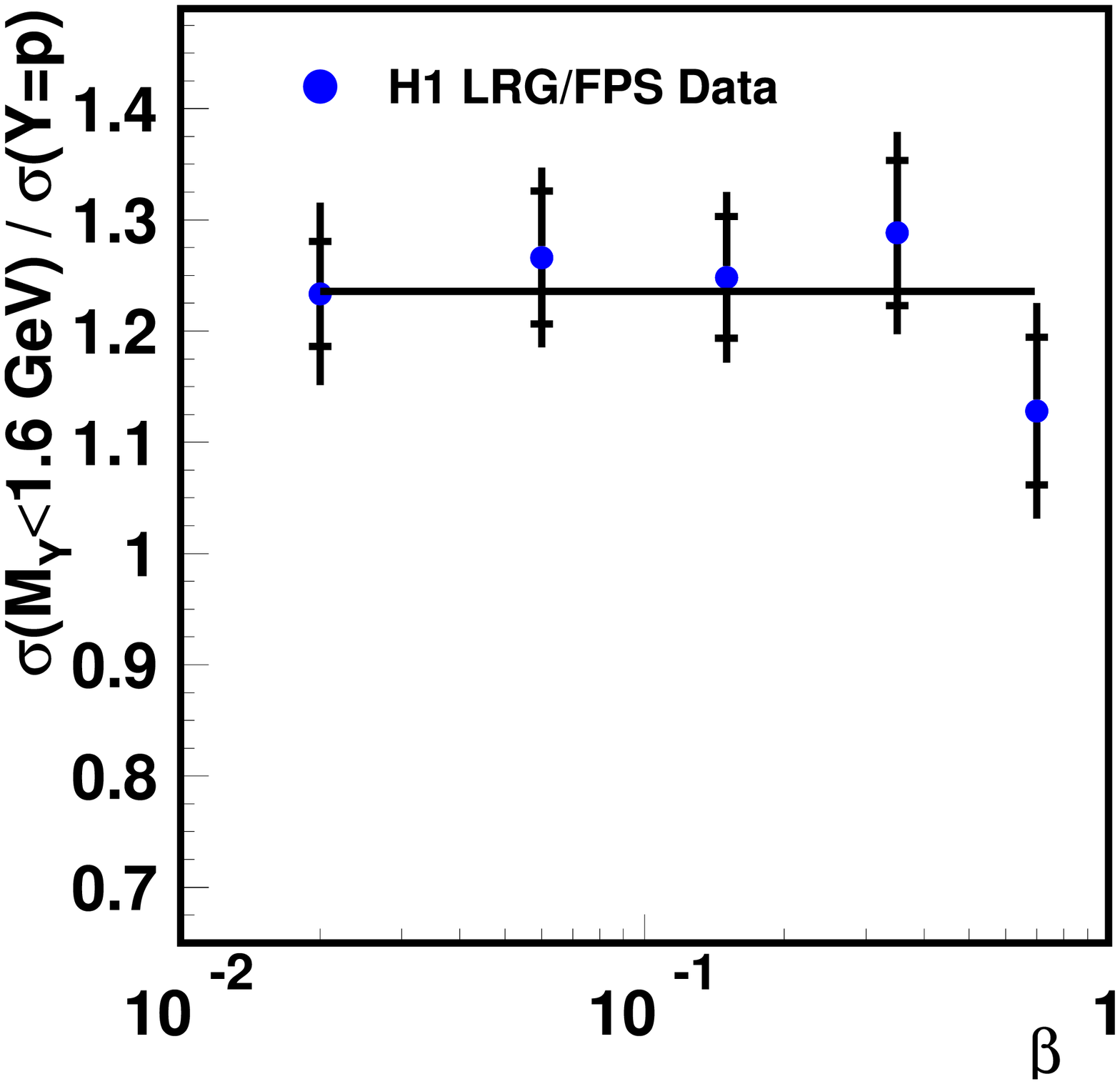,width=0.5\linewidth}}
    \put(40,-3){\epsfig{file=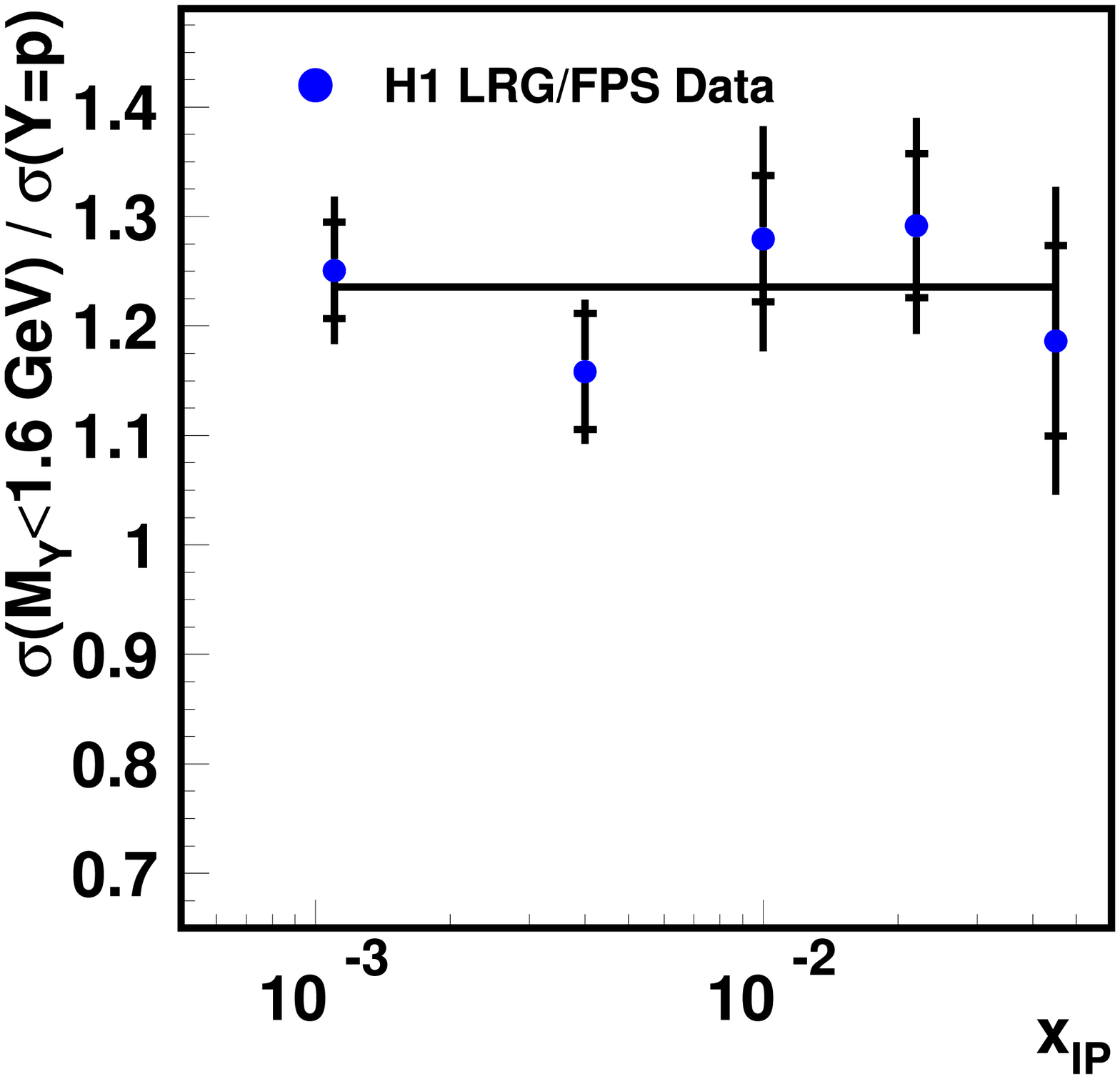,width=0.5\linewidth}}
    \put(40,177){\bf{\Large{(a)}}}
    \put(120,177){\bf{\Large{(b)}}}
    \put(80,80){\bf{\Large{(c)}}}
 \end{picture}
 \end{center}
 \caption{The ratio of the diffractive cross section for $M_Y < 1.6 \ {\rm GeV}$ and
$|t| < 1 \ {\rm GeV^2}$ to that for $Y = p$ and $|t| < 1 \ {\rm GeV^2}$,
obtained from $\sigma_r^{D(3)}$ 
measurements using the LRG and FPS methods. The results are shown
as a function of (a) $Q^2$, (b) $\beta$ and (c) $\xpom$, after
averaging over the other variables.
The lines represent the result of a fit to the data assuming no
dependence on any of these variables.
The inner error bars represent the statistical errors.
The outer error bars indicate the statistical and systematic errors added in 
quadrature. Normalisation uncertainties of $12.7 \%$ are not shown.}
\label{fig:LRGratio}
\end{figure}

The ratio of the LRG to the FPS cross section is plotted in 
figure~\ref{fig:LRGratio} as a function of 
$Q^2$, $\beta$ and $\xpom$.
The combined normalisation errors of $12.7\%$ are not shown. Within the 
remaining uncertainties of typically $10\%$ per data point, 
there is no significant dependence on $\beta$, $Q^2$ or $\xpom$. 
The ratio of overall normalisations, LRG / FPS, is
\begin{eqnarray} 
\frac{\sigma (M_Y < 1.6 \ {\rm GeV})}{\sigma (Y = p)} =  
1.23 \ \pm \ 0.03 \ {\rm (stat.)} \ \pm \ 0.16 \ {\rm (syst.)} \ , 
  \label{myfrac}
\end{eqnarray}
the dominant
uncertainties arising from the normalisations of the 
FPS and LRG data. 
This result is consistent with the prediction 
of $1.15 _{-0.08}^{+0.15}$ from 
the DIFFVM generator, where the total
proton-elastic and proton dissociation cross sections are taken to
be equal by default and their ratio is varied
in the range 1:2 to 2:1 for the uncertainties \cite{H1LRG,DIFFVM}. 

Since the FPS measurement extends to larger $\xpom$ values than the LRG
measurement, the FPS data provide complementary constraints on the
sub-leading exchange trajectory. The value of $n_\reg$ obtained in
section~\ref{f2d4sec} is compared with the 
similarly defined parameter obtained 
in \cite{H1LRG} after dividing the latter by the factor 
$1.23 \pm 0.16$ (equation~\ref{myfrac})
to account for the different $M_Y$ ranges of the two measurements. 
Since all other parameters describing the sub-leading trajectory are fixed
to the same values in the two analyses, the ratio of $n_\reg$ results
is then equivalent to the ratio of sub-leading exchange contributions
in the two cross section measurements. 
The dominant model dependence uncertainties largely cancel
when forming this ratio, which is
\begin{eqnarray} 
\frac{\sigma_\reg \ ({\rm LRG})}{\sigma_\reg \ ({\rm FPS})} =  
1.39 \ \pm \ 0.48 \ {\rm (exp.)} \ \pm \ 0.29 \ {\rm (model)} \ ,
\label{mesfrac}
\end{eqnarray}
where the first error is the combined statistical and experimental
systematic uncertainty and the second is the residual model dependence
uncertainty as defined in section~\ref{f2d4sec}.
This result is consistent with unity,
as expected for a dominantly 
isosinglet sub-leading trajectory ($\omega$ or $f$, rather than $\rho$
or $a$ exchanges). It is thus
consistent with the conclusion from charge exchange cross section measurements
obtained by tagging leading neutrons
in DIS at $\xpom = 0.1$, which can be fully attributed to $\pi$
exchange \cite{H1FPS1}. 

\begin{figure}[p] \unitlength 1mm
 \begin{center}
 \begin{picture}(160,180)
    \put(-3,0){\epsfig{file=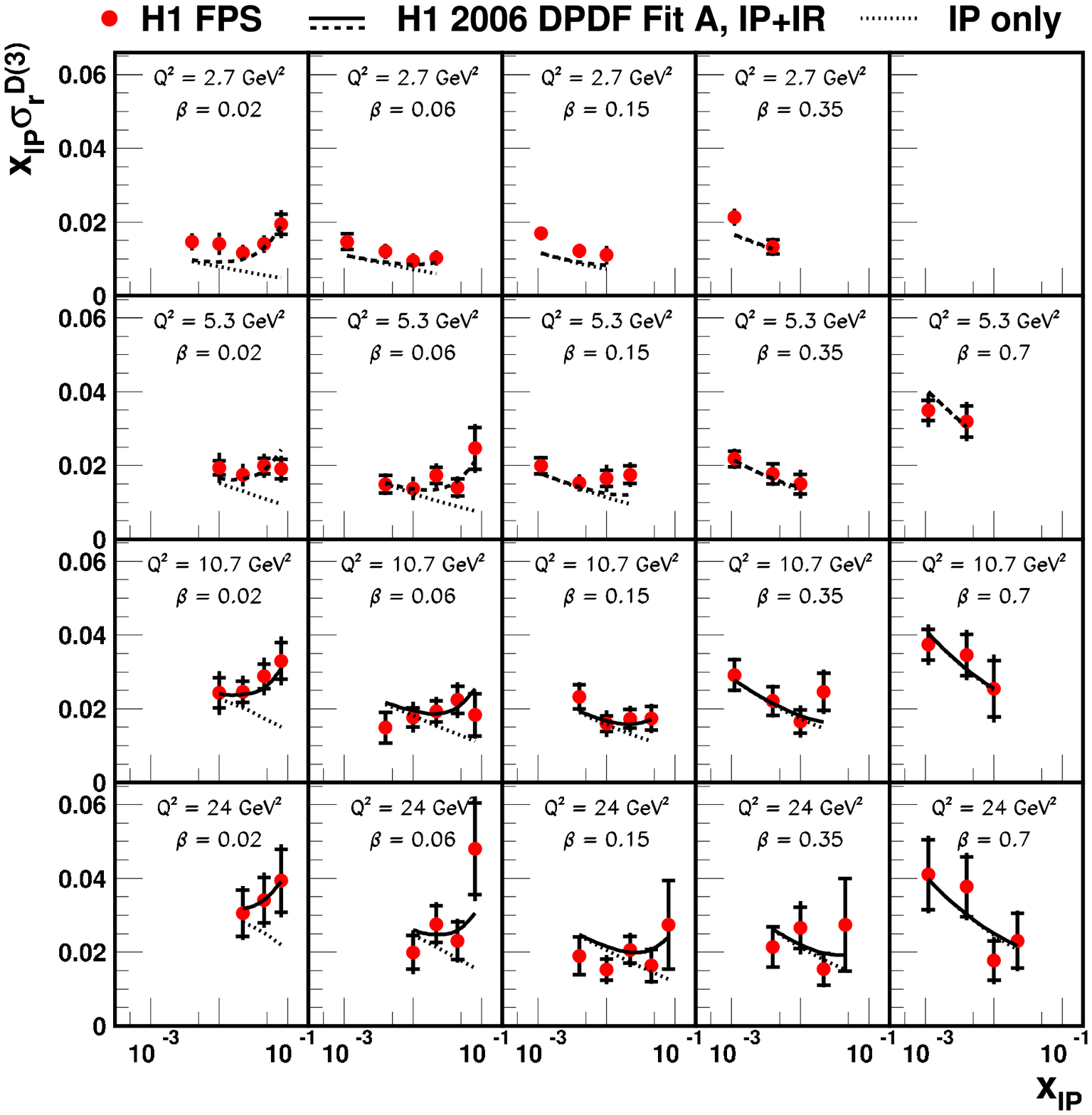,width=\linewidth}}
 \end{picture}
 \end{center}
 \caption{The  diffractive reduced cross section
$\xpom \, \sigma_r^{D(3)}(\beta,Q^2,\xpom)$
for $|t| < 1 \ {\rm GeV^2}$,
shown as a function of $\xpom$ for
different values of $\beta$ and $Q^2$.
The inner error bars represent the statistical errors.
The outer error bars indicate the statistical and systematic 
errors added in quadrature. An overall
normalisation uncertainty of 10.1\% is not shown.
The solid curves represent the results of the `H1 2006 DPDF  
Fit A' to LRG data \cite{H1LRG}, modified as
described in 
section~\ref{lrgcomp}. 
The dashed curves represent the extrapolation 
of this prediction beyond 
the $Q^2$ range which is included in the fit.
The dotted curves indicate the contribution 
of pomeron exchange alone in this model.}
\label{fig:f2d3xp} 
\end{figure}

The predictions of the `2006 DPDF Fit A' to the H1 LRG data \cite{H1LRG} 
are compared with the FPS data in 
figures~\ref{fig:f2d3q2}-\ref{fig:f2d3xp}
after applying a factor of $1/1.39$ (equation~\ref{mesfrac})
to the sub-leading 
exchange contribution in the fit
and an overall normalisation factor of $1 / 1.23$ (equation~\ref{myfrac})
to account for the 
absence of the proton dissociation contribution in the FPS case. 
The FPS data are then well described in the 
region covered by the fit to the LRG data 
($Q^2 \geq 8.5 \ {\rm GeV^2}$). Extrapolating to lower $Q^2$, the 
description remains reasonable.

The good agreement, after accounting for proton dissociation, 
between the LRG 
and the FPS data confirms that the two measurement methods lead
to compatible results, despite 
having very different systematics. 
The lack of any kinematic dependence of the ratio of the two cross
sections shows, within uncertainties, that proton dissociation
with $M_Y < 1.6 \ {\rm GeV}$ can be treated similarly to the
elastic proton case. This supports the factorisation, for both the
pomeron and the sub-leading exchange contributions, of  
processes occuring at the proton vertex from those describing the hard
interaction,
in terms of 
$M_Y$ as well as $t$ (section~\ref{tdep})
and $\xpom$ (section~\ref{f2d4sec}). It also confirms 
that contributions from proton dissociation 
in the LRG measurement do not significantly alter 
the measured $\beta$, $Q^2$ or $\xpom$ dependences and hence cannot
have a large influence on the diffractive gluon density or other information
extracted from the LRG data. 
 
\subsection{Comparison with ZEUS leading proton data.}

\begin{figure}[p] \unitlength 1mm
 \begin{center}
 \begin{picture}(160,180)
    \put(-3,0){\epsfig{file=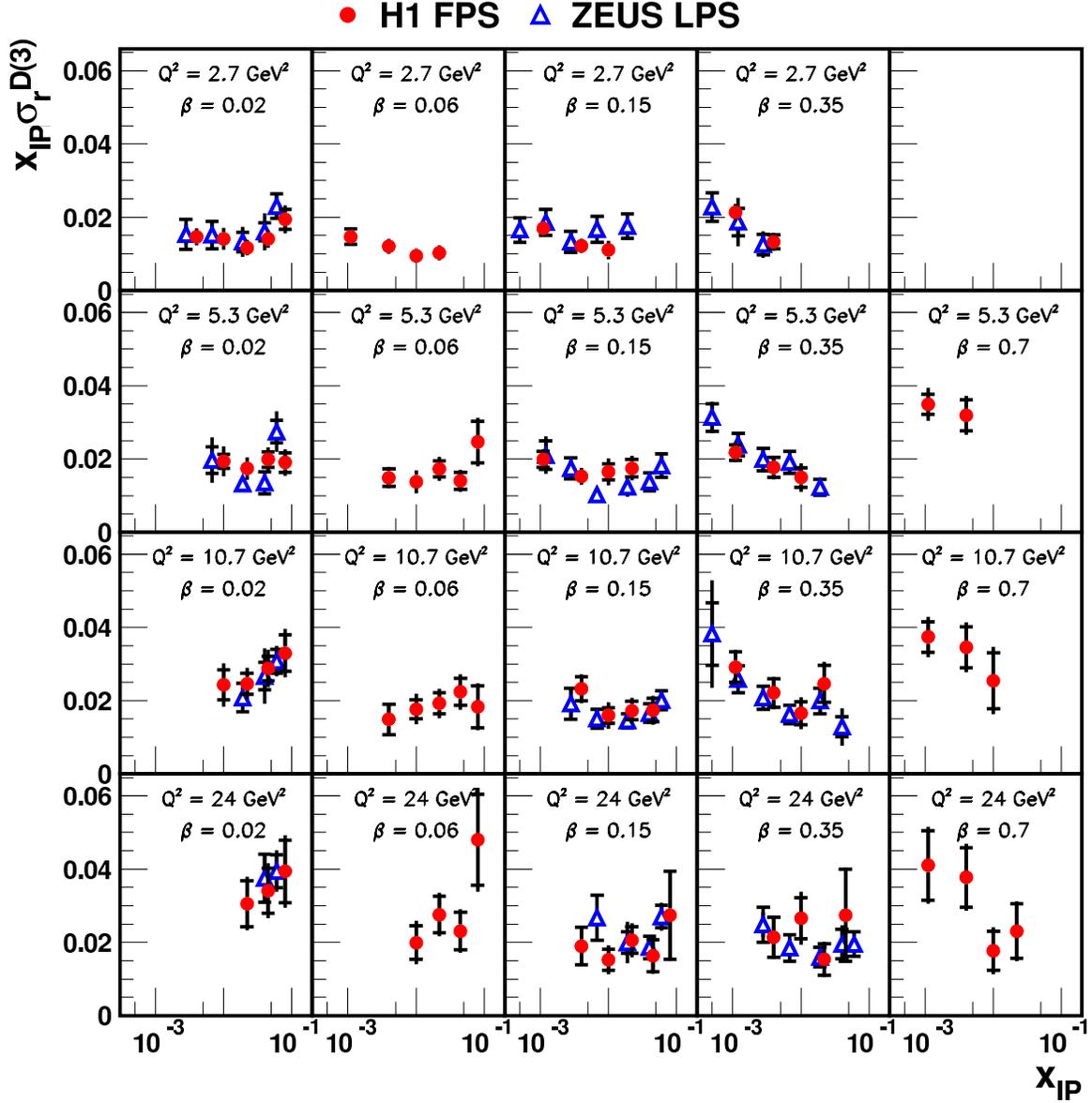,width=\linewidth}}
 \end{picture}
 \end{center}
 \caption{The  diffractive
reduced cross section 
$\xpom \sigma_r^{D(3)}(\beta,Q^2,\xpom)$  
for $|t| < 1 \ {\rm GeV^2}$, shown
as a function of $\xpom$ for
different values of $\beta$ and $Q^2$. H1 FPS data are compared
with ZEUS LPS results \cite{ZEUSLPS}.
The inner error bars represent the statistical errors.
The outer error bars indicate the statistical and systematic errors added in quadrature. Normalisation uncertainties of 
around $10\%$ on each data set are not
shown.}
\label{fig:f2d3zeus} 
\end{figure}

In figure~\ref{fig:f2d3zeus} the FPS $\sigma_r^{D(3)}$ results are 
compared with those of the 
ZEUS collaboration, measured using their
Leading Proton Spectrometer (LPS)~\cite{ZEUSLPS} and also 
integrated over $|t| < 1 \ {\rm GeV^2}$.
The ZEUS data points are interpolated to the 
$\beta$ and $Q^2$ values of this measurement using the
dependences measured in \cite{ZEUSLPS}.
There is very good agreement between the two data sets. 
The ratio of the ZEUS LPS to the H1 FPS data  
averaged over the measured 
kinematic range is 
$~0.92 \pm {\rm 0.04(stat.)} \pm {\rm 0.03(syst.)} \pm 0.15 {\rm(norm.)}$, 
which is 
consistent with unity taking into account the 
dominant normalisation uncertainties. 
Within the errors, there is no 
$\xpom$, $\beta$ or $Q^2$ dependence of the ratio. The sub-leading exchange 
contributes at a similar level at high $\xpom$ and low $\beta$
in both data sets. 

\section{Summary}

A semi-inclusive cross section measurement is presented
for the diffractive deep-inelastic scattering 
process $ep \rightarrow eXp$. 
The results are obtained using data taken with the H1 detector at HERA,
where the scattered
proton carries at least $90\%$ 
of the incoming proton momentum and is measured in the Forward Proton 
Spectrometer (FPS). 
The FPS data are in good agreement with those of the ZEUS collaboration 
obtained with their Leading Proton 
Spectrometer.

The $t$-dependence is parameterised by an exponential
function such that ${\rm d} \sigma / {\rm d} t \propto e^{Bt}$. 
The resulting values of the slope parameter $B$ 
in the pomeron dominated range,
$\xpom \leq 0.0094$, are close to $6 \ {\rm GeV^{-2}}$ and are 
independent of $\xpom$ in this range within errors, favouring an
effective pomeron trajectory slope
$\alpha_\pom^\prime$ which is close to zero. 
There is also no significant $Q^2$ or $\beta$ dependence of $B$.
The slope parameter decreases 
to around $4 \ {\rm GeV^{-2}}$ in the 
higher $\xpom$ region, where an additional 
sub-leading exchange is found to contribute. 

The diffractive reduced cross section 
$\sigma_r^{D(4)}(\beta,Q^2,\xpom,t)$ is measured at $|t|=0.25~\GeV^2$. The 
$\xpom$ dependence is described using 
a model which is motivated by Regge phenomenology, in which a leading 
pomeron and a
sub-leading exchange contribute. The effective pomeron intercept describing the data is 
$\alpha_{\pom}(0)=1.114 \pm 0.018 \ ({\rm stat.}) 
\pm 0.012 \ ({\rm syst.}) \,
^{+0.040}_{-0.020} \ ({\rm model})$. 

The data are also analysed in terms of the 
diffractive reduced cross section $\sigma_r^{D(3)}$, obtained by integrating
$\sigma_r^{D(4)}$ over the range $|t_{\rm min}| < |t| < 1~{\GeV}^2$. 
At fixed $\xpom$, a relatively flat $\beta$ dependence is 
observed over most of the kinematic range.
The data display 
scaling violations with positive $\partial \sigma_r^D / \partial \ln Q^2$,
except at the highest values of $\beta \sim 0.7$.

The FPS data are compared with the results of an
H1 measurement using events selected on the basis of a large rapidity 
gap (LRG) rather than a leading proton, which
includes proton dissociation to states with masses 
$M_Y < 1.6 \ {\rm GeV}$.
The ratio of the LRG to the FPS cross section is 
$1.23  \pm 0.03~{\rm (stat.)} \pm 0.16~{\rm (syst.)}$, 
independently of $Q^2$, $\beta$ and $\xpom$ within the uncertainties. 
Apart from this normalisation factor, the FPS and LRG measurements are in
remarkably good agreement, despite having very different sources of
systematic error.
The magnitude of the 
sub-leading exchange 
component in the FPS data is compatible with that obtained 
from the LRG data, suggesting that 
charge exchange contributions in the latter are small. 
Within the present
uncertainties, the H1 diffractive DIS data are thus compatible with
the factorisation of the variables 
$\xpom$, $t$ and $M_Y$ associated with the proton 
vertex from the variables $\beta$ and $Q^2$, which describe the hard 
interaction, holding separately for the pomeron and 
for the sub-leading exchange trajectory.

\section*{Acknowledgements}

We are grateful to the HERA machine group whose outstanding
efforts have made this experiment possible.
We thank the engineers and technicians for their work in constructing and
maintaining the H1 detector, our funding agencies for
financial support, the DESY technical staff for continual assistance
and the DESY directorate for support and for the
hospitality which they extend to the non-DESY members of the 
collaboration.



\newpage
\renewcommand{\arraystretch}{1.5}
\begin{table}[ht]
\centering
  \begin{tabular}{| c | c | c | c | c | c | c |}
\hline
  $Q^2~{\rm bin} \ [\GeV^2]$ & $\langle Q^2 \rangle \ [\GeV^2]$ & $\beta~{\rm bin}$ & $\langle \beta \rangle$ &
  $\xpom~{\rm bin}$ & $\langle \xpom \rangle$ & $B~[\GeV^{-2}]$ \\
\hline
  {~2~-~50} & {~5.4} & {0.004~-~1} & {0.4~~} & {0.0002~-~0.002} & {0.0009} & $6.21 \pm 0.46^{+0.75}_{-0.35}$ \\
  {~2~-~50} & {~7.5} & {0.004~-~1} & {0.23~} & {0.002~~-~0.006} & {0.0036} & $6.26 \pm 0.59^{+0.62}_{-0.25}$ \\
  {~2~-~50} & {~7.9} & {0.004~-~1} & {0.1~~} & {0.006~~-~0.014} & {0.0094} & $6.14 \pm 0.44^{+0.58}_{-0.22}$ \\
  {~2~-~50} & {~9.0} & {0.004~-~1} & {0.06~} & {0.014~~-~0.03~} & {0.021~} & $5.36 \pm 0.53^{+0.66}_{-0.28}$ \\
  {~2~-~50} & {10.3} & {0.004~-~1} & {0.037} & {0.03~~~-~0.06~} & {0.042~} & $4.16 \pm 0.50^{+0.61}_{-0.26}$ \\
  {~2~-~50} & {12.1} & {0.004~-~1} & {0.023} & {0.06~~~-~0.1~~} & {0.076~} & $4.48 \pm 0.56^{+0.33}_{-0.07}$ \\
\hline
\end{tabular}
 \caption{
The slope parameter $B$, extracted from fits
to the data of the form  ${\rm d}\sigma / {\rm d}t \propto e^{B t}$ 
in different regions of $\xpom$. 
The mean values of $Q^2$, $\beta$ and $\xpom$ are 
also shown for each measurement. 
The first uncertainty given is statistical, the second systematic.}
 \label{table:tdep}
\end{table}

\vspace*{1cm}

\begin{table}[h]
\centering
  \begin{tabular}{| c | c | c | c | c | c | c |}
\hline
  $Q^2~{\rm bin} \ [\GeV^2]$ & $\langle Q^2 \rangle \ [\GeV^2]$ & $\beta~{\rm bin}$ & $\langle \beta \rangle$ &
  $\xpom~{\rm bin}$ & $\langle \xpom \rangle$ & $B~[\GeV^{-2}]$ \\
\hline
  {~2~-~50} & {~5.1} & {0.004~-~0.04} & {0.019} & {0.0002~-~0.03} & {0.013~} & $6.41 \pm 0.58^{+0.85}_{-0.48}$ \\
  {~2~-~50} & {~9.1} & {0.004~-~0.04} & {0.015} & {0.03~~~-~0.1~} & {0.054~} & $4.14 \pm 0.43^{+0.72}_{-0.51}$ \\
  {~2~-~50} & {~7.9} & {0.04~~-~0.25} & {0.12~} & {0.0002~-~0.03} & {0.0074} & $5.60 \pm 0.40^{+0.71}_{-0.41}$ \\
  {~2~-~50} & {16.3} & {0.04~~-~0.25} & {0.082} & {0.03~~~-~0.1~} & {0.048~} & $4.41 \pm 0.82^{+0.66}_{-0.51}$ \\
  {~2~-~50} & {~8.4} & {0.25~~-~1~~~} & {0.51~} & {0.0002~-~0.03} & {0.0027} & $6.73 \pm 0.41^{+0.67}_{-0.39}$ \\
  {~2~-~~4} & {~2.9} & {0.004~-~1~~~} & {0.19~} & {0.0002~-~0.03} & {0.0065} & $5.78 \pm 0.39^{+0.83}_{-0.47}$ \\
  {~2~-~~4} & {~3.0} & {0.004~-~1~~~} & {0.016} & {0.03~~~-~0.1~} & {0.051~} & $5.42 \pm 0.87^{+0.73}_{-0.57}$ \\
  {~4~-~10} & {~6.2} & {0.004~-~1~~~} & {0.23~} & {0.0002~-~0.03} & {0.0077} & $6.72 \pm 0.40^{+0.70}_{-0.39}$ \\
  {~4~-~10} & {~6.6} & {0.004~-~1~~~} & {0.024} & {0.03~~~-~0.1~} & {0.052~} & $4.13 \pm 0.60^{+0.66}_{-0.49}$ \\
  {10~-~50} & {18.8} & {0.004~-~1~~~} & {0.26~} & {0.0002~-~0.03} & {0.01~~} & $5.96 \pm 0.68^{+0.72}_{-0.44}$ \\
  {10~-~50} & {21.2} & {0.004~-~1~~~} & {0.054} & {0.03~~~-~0.1~} & {0.055~} & $3.62 \pm 0.65^{+0.43}_{-0.44}$ \\
\hline
\end{tabular}
 \caption{
The slope parameter $B$ extracted from fits
to the data of the form  ${\rm d}\sigma / {\rm d}t \propto e^{B t}$ 
in different regions of $\xpom$, $\beta$ and $Q^2$. 
The mean values of these kinematic variables are 
also given for each measurement. 
The first uncertainty given is statistical, the second systematic.}
 \label{table:tdep2}
\end{table}

\newpage
\begin{table}[ht]
\centering
  \begin{tabular}{|c | c | c | c | c | c|}
\hline 
  $Q^2~[\GeV^2]$ & $\beta$ & $\xpom$ & $\xpom \sigma_r^{D(4)}~[\GeV^{-2}]$ & $\xpom \sigma_r^{D(3)}$ \\       
\hline 
  2.7 &  0.02 &  0.0040 & $0.0194 \pm 0.0023^{+0.0027}_{-0.0013}$ & $0.0147 \pm 0.0017^{+0.0021}_{-0.0010}$ \\
  2.7 &  0.02 &  0.0100 & $0.0193 \pm 0.0016^{+0.0033}_{-0.0041}$ & $0.0141 \pm 0.0012^{+0.0025}_{-0.0030}$ \\
  2.7 &  0.02 &  0.0220 & $0.0163 \pm 0.0016^{+0.0021}_{-0.0028}$ & $0.0116 \pm 0.0011^{+0.0016}_{-0.0020}$ \\
  2.7 &  0.02 &  0.0450 & $0.0209 \pm 0.0022^{+0.0025}_{-0.0031}$ & $0.0140 \pm 0.0015^{+0.0018}_{-0.0021}$ \\
  2.7 &  0.02 &  0.0800 & $0.0306 \pm 0.0042^{+0.0036}_{-0.0047}$ & $0.0195 \pm 0.0027^{+0.0024}_{-0.0030}$ \\
  2.7 &  0.06 &  0.0011 & $0.0192 \pm 0.0027^{+0.0018}_{-0.0010}$ & $0.0147 \pm 0.0021^{+0.0015}_{-0.0008}$ \\
  2.7 &  0.06 &  0.0040 & $0.0159 \pm 0.0022^{+0.0020}_{-0.0009}$ & $0.0120 \pm 0.0016^{+0.0016}_{-0.0007}$ \\
  2.7 &  0.06 &  0.0100 & $0.0129 \pm 0.0014^{+0.0020}_{-0.0025}$ & $0.0095 \pm 0.0011^{+0.0015}_{-0.0019}$ \\
  2.7 &  0.06 &  0.0220 & $0.0145 \pm 0.0021^{+0.0017}_{-0.0022}$ & $0.0103 \pm 0.0015^{+0.0012}_{-0.0016}$ \\
  2.7 &  0.15 &  0.0011 & $0.0224 \pm 0.0017^{+0.0022}_{-0.0012}$ & $0.0170 \pm 0.0013^{+0.0018}_{-0.0009}$ \\
  2.7 &  0.15 &  0.0040 & $0.0161 \pm 0.0019^{+0.0020}_{-0.0009}$ & $0.0122 \pm 0.0014^{+0.0016}_{-0.0007}$ \\
  2.7 &  0.15 &  0.0100 & $0.0149 \pm 0.0019^{+0.0026}_{-0.0029}$ & $0.0110 \pm 0.0014^{+0.0020}_{-0.0022}$ \\
  2.7 &  0.35 &  0.0011 & $0.0279 \pm 0.0021^{+0.0027}_{-0.0015}$ & $0.0213 \pm 0.0016^{+0.0022}_{-0.0012}$ \\
  2.7 &  0.35 &  0.0040 & $0.0177 \pm 0.0025^{+0.0026}_{-0.0012}$ & $0.0133 \pm 0.0019^{+0.0020}_{-0.0009}$ \\
\hline
  \end{tabular}
 \caption{
The  diffractive reduced cross sections, $\xpom \sigma_r^{D(4)}$ 
measured at $|t|=0.25 \ \GeV^2$, and 
$\xpom \sigma_r^{D(3)}$ integrated over $|t_{\rm min}| < |t| < 1 \ {\GeV}^2$,
measured at $Q^2 = 2.7 \ {\rm GeV^2}$ and various $\beta$ and $\xpom$ values.
The first uncertainty given is statistical, the second systematic.
Normalisation uncertainties of $10.1\%$ are not included.}
\label{table:f2d1}
\end{table}

\newpage

\begin{table}[ht]
\centering
  \begin{tabular}{|c | c | c | c | c | c|}
\hline 
  $Q^2~[\GeV^2]$ & $\beta$ & $\xpom$ & $\xpom \sigma_r^{D(4)}~[\GeV^{-2}]$ & $\xpom \sigma_r^{D(3)}$ \\       
\hline
  5.3 &  0.02 &  0.0100 & $0.0258 \pm 0.0025^{+0.0041}_{-0.0053}$ & $0.0194 \pm 0.0019^{+0.0031}_{-0.0040}$ \\
  5.3 &  0.02 &  0.0220 & $0.0243 \pm 0.0023^{+0.0029}_{-0.0041}$ & $0.0174 \pm 0.0017^{+0.0022}_{-0.0030}$ \\
  5.3 &  0.02 &  0.0450 & $0.0290 \pm 0.0031^{+0.0030}_{-0.0042}$ & $0.0199 \pm 0.0021^{+0.0022}_{-0.0029}$ \\
  5.3 &  0.02 &  0.0800 & $0.0295 \pm 0.0040^{+0.0031}_{-0.0043}$ & $0.0190 \pm 0.0027^{+0.0021}_{-0.0028}$ \\
  5.3 &  0.06 &  0.0040 & $0.0197 \pm 0.0032^{+0.0025}_{-0.0014}$ & $0.0149 \pm 0.0024^{+0.0019}_{-0.0011}$ \\
  5.3 &  0.06 &  0.0100 & $0.0185 \pm 0.0023^{+0.0030}_{-0.0039}$ & $0.0138 \pm 0.0017^{+0.0023}_{-0.0029}$ \\
  5.3 &  0.06 &  0.0220 & $0.0240 \pm 0.0031^{+0.0028}_{-0.0039}$ & $0.0173 \pm 0.0022^{+0.0021}_{-0.0028}$ \\
  5.3 &  0.06 &  0.0450 & $0.0208 \pm 0.0034^{+0.0022}_{-0.0031}$ & $0.0140 \pm 0.0023^{+0.0016}_{-0.0021}$ \\
  5.3 &  0.06 &  0.0800 & $0.0369 \pm 0.0087^{+0.0041}_{-0.0060}$ & $0.0246 \pm 0.0057^{+0.0028}_{-0.0041}$ \\
  5.3 &  0.15 &  0.0011 & $0.0260 \pm 0.0029^{+0.0024}_{-0.0015}$ & $0.0199 \pm 0.0022^{+0.0020}_{-0.0012}$ \\
  5.3 &  0.15 &  0.0040 & $0.0202 \pm 0.0021^{+0.0027}_{-0.0012}$ & $0.0153 \pm 0.0016^{+0.0021}_{-0.0010}$ \\
  5.3 &  0.15 &  0.0100 & $0.0222 \pm 0.0030^{+0.0035}_{-0.0044}$ & $0.0165 \pm 0.0022^{+0.0026}_{-0.0033}$ \\
  5.3 &  0.15 &  0.0220 & $0.0243 \pm 0.0034^{+0.0027}_{-0.0037}$ & $0.0175 \pm 0.0024^{+0.0020}_{-0.0026}$ \\
  5.3 &  0.35 &  0.0011 & $0.0286 \pm 0.0028^{+0.0025}_{-0.0019}$ & $0.0218 \pm 0.0021^{+0.0020}_{-0.0015}$ \\
  5.3 &  0.35 &  0.0040 & $0.0232 \pm 0.0036^{+0.0027}_{-0.0015}$ & $0.0177 \pm 0.0027^{+0.0021}_{-0.0012}$ \\
  5.3 &  0.35 &  0.0100 & $0.0200 \pm 0.0036^{+0.0035}_{-0.0041}$ & $0.0149 \pm 0.0027^{+0.0027}_{-0.0031}$ \\
  5.3 &  0.70 &  0.0011 & $0.0460 \pm 0.0037^{+0.0049}_{-0.0040}$ & $0.0349 \pm 0.0028^{+0.0039}_{-0.0031}$ \\
  5.3 &  0.70 &  0.0040 & $0.0419 \pm 0.0056^{+0.0049}_{-0.0033}$ & $0.0319 \pm 0.0043^{+0.0039}_{-0.0026}$ \\
\hline
  \end{tabular}
 \caption{
The  diffractive reduced cross sections, $\xpom \sigma_r^{D(4)}$ 
measured at $|t|=0.25 \ \GeV^2$, and 
$\xpom \sigma_r^{D(3)}$ integrated over $|t_{\rm min}| < |t| < 1 \ {\GeV}^2$,
measured at $Q^2 = 5.3 \ {\rm GeV^2}$ and various $\beta$ and $\xpom$ values.
The first uncertainty given is statistical, the second systematic.
Normalisation uncertainties of $10.1\%$ are not included.}
\label{table:f2d2}
\end{table}

\newpage

\begin{table}[ht]
\centering
  \begin{tabular}{|c | c | c | c | c | c|}
\hline 
  $Q^2~[\GeV^2]$ & $\beta$ & $\xpom$ & $\xpom \sigma_r^{D(4)}~[\GeV^{-2}]$ & $\xpom \sigma_r^{D(3)}$ \\       
\hline
 10.7 &  0.02 &  0.0100 & $0.0325 \pm 0.0054^{+0.0050}_{-0.0064}$ & $0.0243 \pm 0.0040^{+0.0039}_{-0.0048}$ \\
 10.7 &  0.02 &  0.0220 & $0.0345 \pm 0.0040^{+0.0037}_{-0.0050}$ & $0.0246 \pm 0.0028^{+0.0027}_{-0.0036}$ \\
 10.7 &  0.02 &  0.0450 & $0.0422 \pm 0.0049^{+0.0047}_{-0.0064}$ & $0.0288 \pm 0.0033^{+0.0034}_{-0.0044}$ \\
 10.7 &  0.02 &  0.0800 & $0.0502 \pm 0.0075^{+0.0053}_{-0.0076}$ & $0.0330 \pm 0.0050^{+0.0036}_{-0.0051}$ \\
 10.7 &  0.06 &  0.0040 & $0.0196 \pm 0.0056^{+0.0024}_{-0.0012}$ & $0.0149 \pm 0.0042^{+0.0019}_{-0.0009}$ \\
 10.7 &  0.06 &  0.0100 & $0.0236 \pm 0.0034^{+0.0038}_{-0.0048}$ & $0.0177 \pm 0.0026^{+0.0029}_{-0.0036}$ \\
 10.7 &  0.06 &  0.0220 & $0.0269 \pm 0.0041^{+0.0032}_{-0.0044}$ & $0.0193 \pm 0.0029^{+0.0023}_{-0.0032}$ \\
 10.7 &  0.06 &  0.0450 & $0.0329 \pm 0.0054^{+0.0038}_{-0.0053}$ & $0.0224 \pm 0.0037^{+0.0027}_{-0.0037}$ \\
 10.7 &  0.06 &  0.0800 & $0.0278 \pm 0.0087^{+0.0033}_{-0.0051}$ & $0.0184 \pm 0.0057^{+0.0023}_{-0.0034}$ \\
 10.7 &  0.15 &  0.0040 & $0.0309 \pm 0.0044^{+0.0034}_{-0.0020}$ & $0.0232 \pm 0.0033^{+0.0027}_{-0.0016}$ \\
 10.7 &  0.15 &  0.0100 & $0.0213 \pm 0.0028^{+0.0035}_{-0.0044}$ & $0.0160 \pm 0.0021^{+0.0027}_{-0.0034}$ \\
 10.7 &  0.15 &  0.0220 & $0.0240 \pm 0.0036^{+0.0028}_{-0.0038}$ & $0.0173 \pm 0.0026^{+0.0020}_{-0.0028}$ \\
 10.7 &  0.15 &  0.0450 & $0.0254 \pm 0.0047^{+0.0026}_{-0.0036}$ & $0.0174 \pm 0.0032^{+0.0019}_{-0.0025}$ \\
 10.7 &  0.35 &  0.0011 & $0.0382 \pm 0.0055^{+0.0032}_{-0.0021}$ & $0.0292 \pm 0.0042^{+0.0026}_{-0.0017}$ \\
 10.7 &  0.35 &  0.0040 & $0.0292 \pm 0.0051^{+0.0035}_{-0.0018}$ & $0.0221 \pm 0.0039^{+0.0027}_{-0.0014}$ \\
 10.7 &  0.35 &  0.0100 & $0.0222 \pm 0.0042^{+0.0034}_{-0.0043}$ & $0.0166 \pm 0.0032^{+0.0026}_{-0.0032}$ \\
 10.7 &  0.35 &  0.0220 & $0.0341 \pm 0.0070^{+0.0038}_{-0.0049}$ & $0.0246 \pm 0.0051^{+0.0028}_{-0.0036}$ \\
 10.7 &  0.70 &  0.0011 & $0.0492 \pm 0.0055^{+0.0047}_{-0.0044}$ & $0.0374 \pm 0.0041^{+0.0038}_{-0.0034}$ \\
 10.7 &  0.70 &  0.0040 & $0.0454 \pm 0.0073^{+0.0061}_{-0.0040}$ & $0.0346 \pm 0.0056^{+0.0048}_{-0.0031}$ \\
 10.7 &  0.70 &  0.0100 & $0.0339 \pm 0.0102^{+0.0064}_{-0.0071}$ & $0.0254 \pm 0.0077^{+0.0049}_{-0.0053}$ \\
\hline
  \end{tabular}
 \caption{
The  diffractive reduced cross sections, $\xpom \sigma_r^{D(4)}$ 
measured at $|t|=0.25 \ \GeV^2$, and 
$\xpom \sigma_r^{D(3)}$ integrated over $|t_{\rm min}| < |t| < 1 \ {\GeV}^2$,
measured at $Q^2 = 10.7 \ {\rm GeV^2}$ and various $\beta$ and $\xpom$ values.
The first uncertainty given is statistical, the second systematic.
Normalisation uncertainties of $10.1\%$ are not included.}
\label{table:f2d3}
\end{table}

\newpage

\begin{table}[ht]
\centering
  \begin{tabular}{|c | c | c | c | c | c|}
\hline 
  $Q^2~[\GeV^2]$ & $\beta$ & $\xpom$ & $\xpom \sigma_r^{D(4)}~[\GeV^{-2}]$ & $\xpom \sigma_r^{D(3)}$ \\       
\hline
 24.0 &  0.02 &  0.0220 & $0.0425 \pm 0.0088^{+0.0044}_{-0.0057}$ & $0.0306 \pm 0.0063^{+0.0033}_{-0.0042}$ \\
 24.0 &  0.02 &  0.0450 & $0.0497 \pm 0.0090^{+0.0052}_{-0.0069}$ & $0.0341 \pm 0.0062^{+0.0038}_{-0.0048}$ \\
 24.0 &  0.02 &  0.0800 & $0.0596 \pm 0.0128^{+0.0062}_{-0.0086}$ & $0.0394 \pm 0.0085^{+0.0043}_{-0.0058}$ \\
 24.0 &  0.06 &  0.0100 & $0.0264 \pm 0.0061^{+0.0043}_{-0.0055}$ & $0.0200 \pm 0.0046^{+0.0033}_{-0.0042}$ \\
 24.0 &  0.06 &  0.0220 & $0.0386 \pm 0.0070^{+0.0038}_{-0.0048}$ & $0.0276 \pm 0.0050^{+0.0028}_{-0.0034}$ \\
 24.0 &  0.06 &  0.0450 & $0.0334 \pm 0.0075^{+0.0035}_{-0.0047}$ & $0.0231 \pm 0.0052^{+0.0026}_{-0.0033}$ \\
 24.0 &  0.06 &  0.0800 & $0.0740 \pm 0.0187^{+0.0081}_{-0.0120}$ & $0.0480 \pm 0.0124^{+0.0055}_{-0.0079}$ \\
 24.0 &  0.15 &  0.0040 & $0.0252 \pm 0.0068^{+0.0027}_{-0.0013}$ & $0.0190 \pm 0.0052^{+0.0021}_{-0.0010}$ \\
 24.0 &  0.15 &  0.0100 & $0.0204 \pm 0.0038^{+0.0031}_{-0.0040}$ & $0.0152 \pm 0.0028^{+0.0024}_{-0.0030}$ \\
 24.0 &  0.15 &  0.0220 & $0.0287 \pm 0.0051^{+0.0033}_{-0.0044}$ & $0.0206 \pm 0.0036^{+0.0024}_{-0.0032}$ \\
 24.0 &  0.15 &  0.0450 & $0.0240 \pm 0.0064^{+0.0029}_{-0.0039}$ & $0.0164 \pm 0.0044^{+0.0021}_{-0.0027}$ \\
 24.0 &  0.15 &  0.0800 & $0.0414 \pm 0.0181^{+0.0039}_{-0.0048}$ & $0.0274 \pm 0.0120^{+0.0028}_{-0.0033}$ \\
 24.0 &  0.35 &  0.0040 & $0.0281 \pm 0.0072^{+0.0031}_{-0.0016}$ & $0.0214 \pm 0.0055^{+0.0025}_{-0.0013}$ \\
 24.0 &  0.35 &  0.0100 & $0.0356 \pm 0.0074^{+0.0054}_{-0.0068}$ & $0.0266 \pm 0.0056^{+0.0041}_{-0.0051}$ \\
 24.0 &  0.35 &  0.0220 & $0.0210 \pm 0.0060^{+0.0024}_{-0.0032}$ & $0.0153 \pm 0.0043^{+0.0018}_{-0.0024}$ \\
 24.0 &  0.35 &  0.0450 & $0.0400 \pm 0.0183^{+0.0035}_{-0.0048}$ & $0.0274 \pm 0.0126^{+0.0026}_{-0.0034}$ \\
 24.0 &  0.70 &  0.0011 & $0.0535 \pm 0.0125^{+0.0066}_{-0.0044}$ & $0.0410 \pm 0.0095^{+0.0052}_{-0.0035}$ \\
 24.0 &  0.70 &  0.0040 & $0.0494 \pm 0.0106^{+0.0063}_{-0.0042}$ & $0.0378 \pm 0.0081^{+0.0049}_{-0.0033}$ \\
 24.0 &  0.70 &  0.0100 & $0.0238 \pm 0.0071^{+0.0040}_{-0.0050}$ & $0.0177 \pm 0.0054^{+0.0031}_{-0.0038}$ \\
 24.0 &  0.70 &  0.0220 & $0.0315 \pm 0.0103^{+0.0039}_{-0.0053}$ & $0.0231 \pm 0.0074^{+0.0029}_{-0.0039}$ \\
\hline
\end{tabular}
 \caption{
The  diffractive reduced cross sections, $\xpom \sigma_r^{D(4)}$ 
measured at $|t|=0.25 \ \GeV^2$, and 
$\xpom \sigma_r^{D(3)}$ integrated over $|t_{\rm min}| < |t| < 1 \ {\GeV}^2$,
measured at $Q^2 = 24 \ {\rm GeV^2}$ and various $\beta$ and $\xpom$ values.
The first uncertainty given is statistical, the second systematic.
Normalisation uncertainties of $10.1\%$ are not included.}
\label{table:f2d4}
\end{table}

\end{document}